\documentclass[prb, twocolumn,url,nofootinbib]{revtex4}
\usepackage{amsmath,bm}
\usepackage{graphicx,pdflscape}

\usepackage{hyperref}
\usepackage{color,subfigure}
\usepackage{placeins}

\newcommand{\beq}{\begin{eqnarray}}
\newcommand{\eeq}{\end{eqnarray}}
\newcommand{\barray}{\begin{eqnarray}}
\newcommand{\earray}{\end{eqnarray}}
\newcommand{\disp}[1]{Eq.~(\ref{#1})}

\newcommand{\refdisp}[1]{Ref.~[\onlinecite{#1}]}
\newcommand{\figdisp}[1]{Fig.~(\ref{#1})}

\newcommand{\si}{\sigma}

\newcommand{\tJ}{\ $t$-$J$ \ }
\newcommand{\nn}{\nonumber}

\renewcommand{\emph}{\textit}

\newcommand{\half}{\frac{1}{2}}
\usepackage{atveryend}
\makeatletter
\let\origcitation\citation
\AtEndDocument{\def\mycites{}%
  \def\citation#1{\g@addto@macro\mycites{#1^^J}\origcitation{#1}}}
\AtVeryEndDocument{\newwrite\citeout\immediate\openout\citeout=\jobname.cit
  \immediate\write\citeout{\mycites}\immediate\closeout\citeout}
\makeatother

\begin{document}

\title{Nonresonant Raman  scattering in  extremely correlated Fermi liquids}
\author{Peizhi Mai  and B. Sriram Shastry\\
Physics Department, University of California, Santa Cruz, CA 95064 \\
}
\date{\today}

\begin{abstract}
We present theoretical results for the optical conductivity and the nonresonant Raman susceptibilities for three principal polarization geometries relevant  to the square lattice. The susceptibilities are obtained  using the recently developed extremely correlated Fermi liquid theory for the two-dimensional $t$-$t'$-$J$ model, where $t$ and $t'$ are the nearest and second neighbor hopping.  Our results sensitively depend on $t$, $t'$. By  studying this quartet of  related  dynamical susceptibilities, and their dependence on $t$, $t'$, doping and temperature,  we  provide a useful framework  of interpreting  and planning future Raman experiments on strongly correlated matter.
\end{abstract}

\maketitle

\section{ Introduction}

Inelastic or Raman scattering of electrons by photons ($e$-$\gamma$)   in strongly correlated systems  is of considerable current  interest. The scattering intensity, given by the Kramers-Heisenberg formula \cite{Kramers-Heisenberg},  consists of a  resonant and a nonresonant piece.  The nonresonant piece  depends only on the energy transfer. In contrast, the resonant piece also depends on the incident energy, and it is the focus of this  work. In typical weakly correlated metals, this  contribution is confined to a small energy window   of a few $m$eV \cite{Wolff-1,Wolff-2}.  Raman scattering theory, if based solely on density fluctuations,  would give a vanishing contribution as $q\to0$ due to the conservation law in that limit. The  early  works of  \refdisp{Wolff-2}  and  \refdisp{Abrikosov-Genkin}  showed that non-parabolic bands lead to the coupling of light to a non-conserved operator (the stress tensor operators discussed below), rather than the density. These operators are exempt from conservation laws that govern the density,  and therefore they can lead to nonresonant Raman scattering.

Recent experiments \cite{Sugai-1,Sugai-2,Sugai-3,Hackl,Qazilbash,Koitzsch,Sauer-Girsh,Klein-Dierker,klein1,Lance1,Lance2,Lance3,Lance4}
in strongly correlated metallic systems, such as High Tc superconductors have added further  complexity to  challenge to our understanding. It is found that 
 the scattering is  $q$-independent and extends over a much larger  energy range ${\cal O}$(eV), and  is also observed to have  a complex $T$ dependence\cite{Sugai-1,Sugai-2,Sugai-3,Lance1,Koitzsch}. To explain these  a systematic  reformulation of light scattering in narrow band systems was developed in \refdisp{raman1,raman2,Freericks,Medici,devereaux,Zawadowski}.    Shastry and Shraiman (SS) \refdisp{raman1,raman2} developed a theory of Raman scattering in Mott-Hubbard systems
 using the Hubbard model, where nonparabolicity of bands is built in correctly, so that the conservation law concerns are taken care of. However  the  large energy spread of the  nonresonant signals remains unaccounted for. It cannot arise from quasi-particles in Fermi liquids, and hence   SS argued that a large contribution from the  incoherent background of the electron spectral function is required to explain the data (see e.g. \cite{Sugai-1,Sugai-2}).   This qualitative  argument is not fine enough to   explain or predict differences in  backgrounds  in different geometries. The latter   remains an unresolved  problem, and it is the focus  of the present  work.

  Progress towards a solution at the microscopic level has been slow since  a suitable   theory  in 2-dimensions displaying such a phenomenon  has been lacking so far. In this work we apply the recently developed extremely correlated Fermi liquid theory (ECFL) \cite{ECFL,sriram-edward} to calculate the Raman cross sections using  the $k$-dependent bare vertices of \refdisp{raman1,raman2}. This theory provides a framework for  controlled calculations  in  the \tJ model, a prototypical model for very strong correlations, and a limiting case of the Hubbard model. The theory has been  successfully benchmarked  against essentially exact   results  in $d=0$\cite{0d},  $d=1$ \cite{1d} as well as $d=\infty$ \cite{infinited}. A recent application of the theory to the physically important case of   $d=2$ in \refdisp{SP,PS} gives detailed results for the spectral functions and the resistivity $\rho$  in the $t$-$t'$-$J$ model, with nearest and second neighboring hopping.  The state obtained in ECFL  at low hole densities has  a very small quasi-particle weight $Z\ll 1$. A significant result is that  
the temperature dependence of resistivity is non-quadratic already at $T\sim 100$K  for low hole doping.

 In this work we apply the  solution found in \refdisp{SP,PS} to
  compute the Raman scattering,  in three standard polarization configuration channels  $A_{1g},B_{1g},B_{2g}$ defined below\cite{Comment-A1g}.   The results are applicable to either electron doping or hole doped cuprates by choosing the sign of $t'$,  and they may apply to other strongly correlated systems as well. Following SS, we also compare the Raman conductivities with the optical conductivity, and we shall focus on the  quartet of these results on various values of material parameters.   

The utility of comparing the  optical conductivity with the Raman response requires a comment. SS\cite{raman1,raman2} suggested that this comparison is useful, since these are exactly related in a limiting situation of $d=\infty$. Further in $d=2,3..$, 
  one often calculates the response  within the bubble diagrams, where again these are related. In the bubble approximation,  also used in the present work, one  evaluates  the current-current and related correlation functions  by retaining only the lowest order $\chi_{JJ}\sim \sum_k (\gamma_k)^2 G(k) G(k)$ (i.e. bubble) terms with  dressed Greens functions and  suitable bare vertices $\gamma$. While this calculation  misses a contribution due to the renormalization of one of the bare vertices $\gamma\to \Gamma$, it is hard to improve on this already difficult calculation for strong correlations, since $G$ is highly non-trivial. An    exception  is the  special case of  $d\to \infty$, where  the vertex corrections vanish. Within the bubble scheme, the  bare Raman and current vertices are different  while  everything else is  the same. Therefore one should be able to relate the two experimental results and explore the differences arising from the bare vertices. The ``pseudo-identity'' of the transport and Raman resistivities have been explored experimentally in \refdisp{Hackl} and finds some support. In this work we use the correct bare vertices in the different geometries to explore the 
various   Raman resistivities to refine the theory. These different bare vertices have a  different dependence on the hopping parameters $t,t'$, and the calculations reflect these  in specific and experimentally testable  ways.

The neglect of vertex corrections also leads to a relationship between various Raman susceptibilities at finite $\omega$.  In the experiments of \refdisp{Sugai-2},  the same quartet of  susceptibilities has been studied and found to have a roughly similar scale for their $\omega$ dependence, although the curve shapes are distinct. On the theoretical side, one interesting aspect of the results of \refdisp{SP,PS} is that the Fermi surface {\em shape} remains very close to that of the non-interacting tight binding model, while of course conserving the area.  Thus the Dyson self energy is a weak function of $\vec{k}$, unlike the strong dependence in 1-dimension \cite{1d}.
This fact implies that the vertex corrections, while nonzero, are modest.

\section{The Raman and current vertices} We use  the $t$-$t'$-$J$ model with a  tight-binding dispersion  \cite{SP} on the square lattice   $\varepsilon(k)= - 2 t 
[\cos(k_x)+\cos(k_y)] - 4 t' \cos(k_x) \cos(k_y)$,
 and we set the lattice constant $a_0\to1$. The  photons modulate the Peierls hopping factors as $t_{ij}\to t_{ij} \exp\{ i e/\hbar \int_i^j  d\vec{r}. \vec{A} \}$, and the second-order expansion coefficients define the scattering operators.
 In this case they  are
 \beq
{\widehat{ \cal J}}_{\alpha,q} = \sum_{k \sigma} {\cal J}_{\alpha}(k) C^\dagger_{k+\half q, \sigma} C_{k- \half q, \si},  
\eeq
where $\alpha$ is a composite index determined by the in-out polarizations of the photon. 
 With that 
the vertices ${\cal J}_\alpha$ for the   three main  Raman channels are 
 \beq
A_{1g}: &&   {\cal J}_{A_{1g}}(k)=  2t (\cos k_x + \cos k_y) + 4 t' \cos k_x \cos k_y, \nn \\
B_{1g}: &&  {\cal J}_{B_{1g}}(k) = 2 t (\cos k_x - \cos k_y), \nn \\
B_{2g}: &&   {\cal J}_{B_{2g}}(k) = -4 t' \sin k_x \sin k_y , \nn \\
\mbox{xx}: &&     {\cal J}_{xx}(k) = 2 \sin k_x (t+2 t' \cos k_y).  \label{vertices}
 \eeq
The  definition of  $\alpha=xx$  corresponds to the particle current along $x$. It integrates the charge current into the same scheme as the Raman scattering.  It is interesting that the $B_{2g}$ vertex is independent of $t$, and is solely governed by $t'$.  The vertex  $B_{1g}$  is complementary given its independence of $t'$.   These geometries sample different parts of $k$ space in interesting ways due to their different $\vec{k}$ dependences.

 We next define the calculated variables, and we display the results for them from  computations  based on the spectral functions found in \refdisp{SP,PS}. Results in the $\omega=0$ DC limit and also at finite $\omega$ are  shown. Finally we discuss the results and their significance.

\section{  Raman and charge susceptibilities}
We  summarize  the formulas for the (nonresonant) Raman  susceptibility, and in the spirit of \refdisp{raman1,raman2}  also define a {\em Raman conductivity and resistivity}  in analogy as follows
\beq
\chi_\alpha(q,z)= \sum_{n m} \frac{p_n-p_m}{\epsilon_m- \epsilon_n -  z}   |\left({\widehat{\cal J}}_{\alpha, q}\right)_{n,m}|^2, \label{chi-def}
\eeq
where $p_n$ is the probability of the state $n$.   For visible light, $q a_0\ll1$ and therefore we set $q\to0$. 
The (nonresonant) Raman intensity ${\cal I}_\alpha$ \cite{Kramers-Heisenberg,Wolff-2,Wolff-1, raman1,raman2} and the Raman conductivities \cite{raman1,raman2} are given by 
\beq
{\cal I}_\alpha(0,\omega)&=& \frac{\chi''_\alpha(0,\omega)}{(1- e^{- \beta  \omega})}, \;\;
 \sigma_\alpha(\omega) =\zeta_\alpha \frac{ \chi''_\alpha(0,\omega) }{N_s \omega}, \label{chidef}
 \eeq
 with $N_s$  the number of sites,  and $\zeta_{xx}=e^2$ accounting for the electric charge in the  conductivity with all other $\zeta_\alpha=1$. In the dc limit we define the Raman resistivities
 \beq
 \rho_\alpha(0) & =& \frac{{N_s}}{\zeta_\alpha} \;  \frac{k_B T}{{\cal I}_\alpha(0,0)} \label{pseudo-1}
\eeq
where for $\alpha=xx$, $\rho_\alpha$ is the usual resistivity.

\begin{figure*}[ht]
\subfigure[{Electrical resistivity $\bar{\rho}_{xx}$}]{\includegraphics[width=.7\columnwidth]{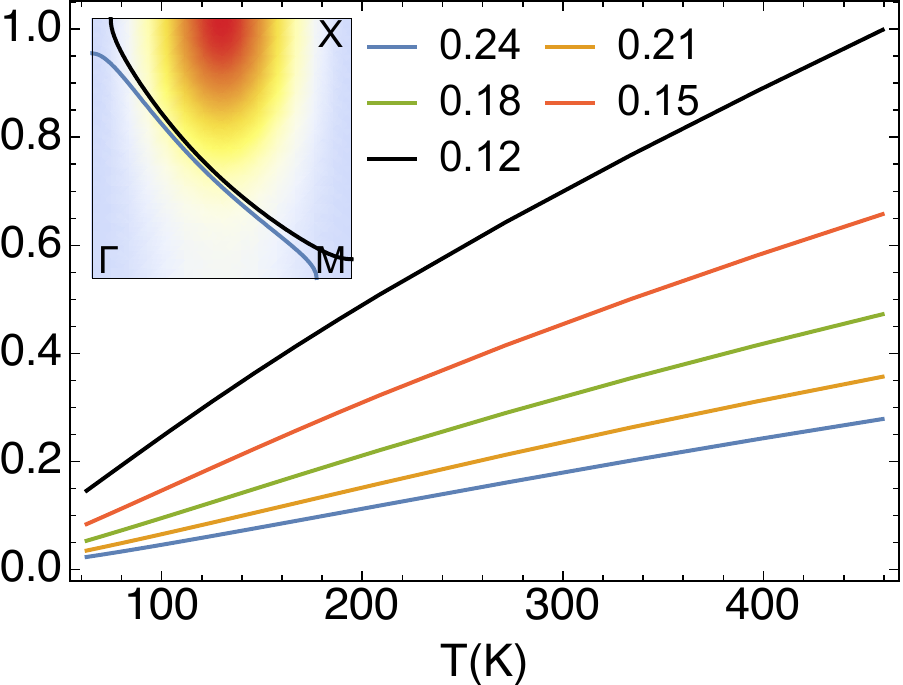}}
\subfigure[{$A_{1g}$ Raman: resistivity $\bar{\rho}_{A_{1g}}$}]{\includegraphics[width=.66\columnwidth]{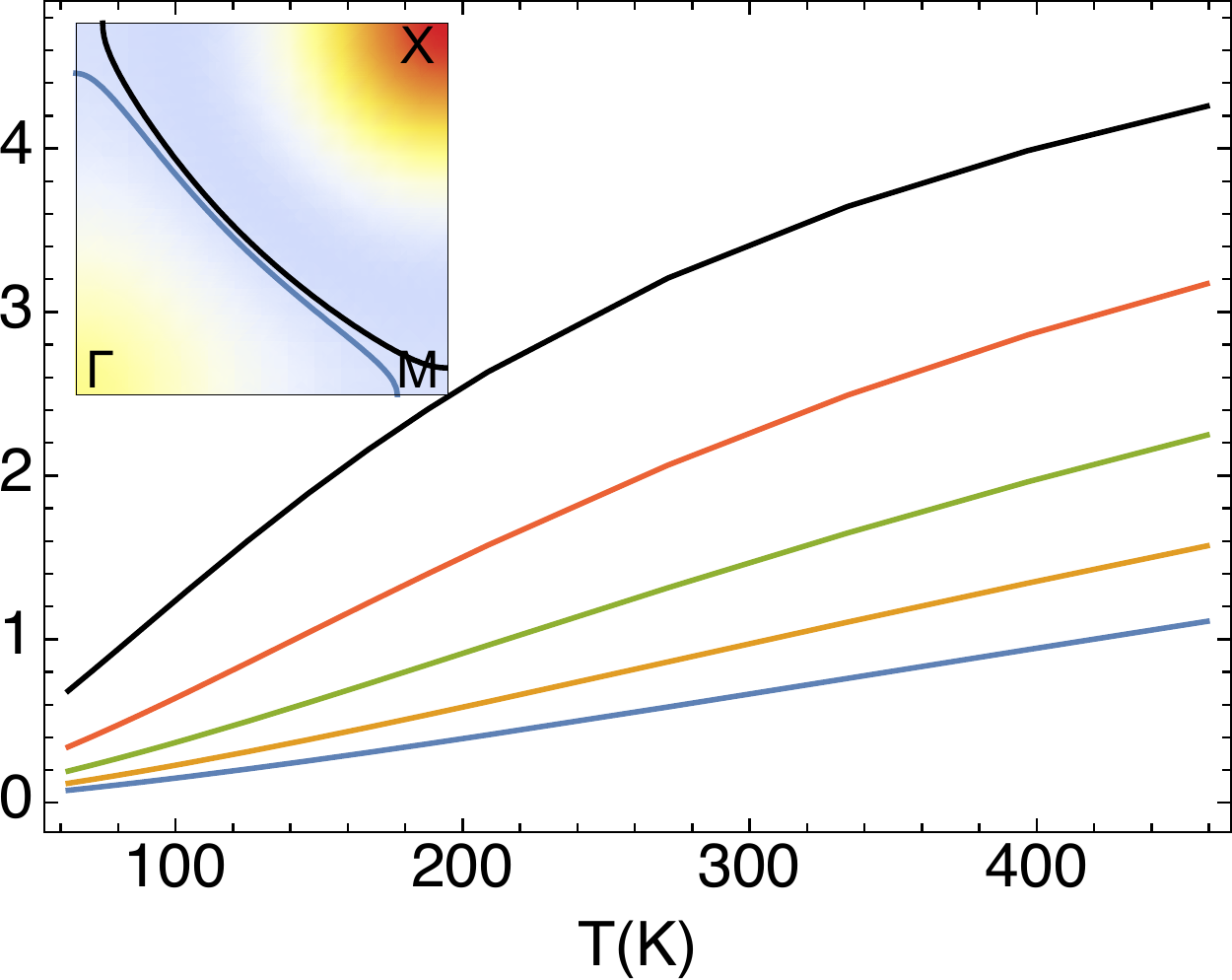}}
 \subfigure[{$B_{1g}$ Raman: resistivity $\bar{\rho}_{B_{1g}}$}]{\includegraphics[width=.7\columnwidth]{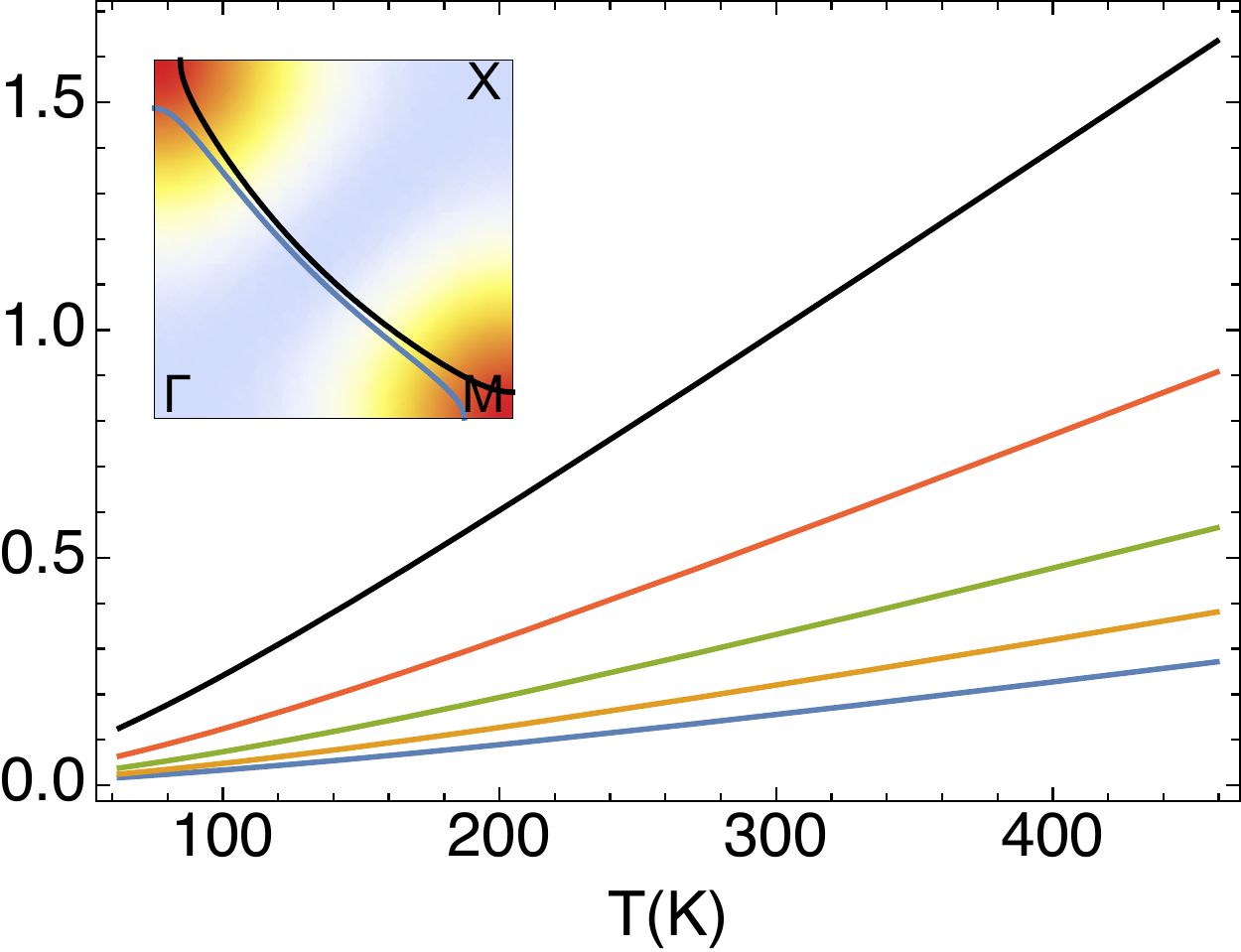}}
\subfigure[{$B_{2g}$ Raman: resistivity $\bar{\rho}_{B_{2g}}$}]{\includegraphics[width=.7\columnwidth]{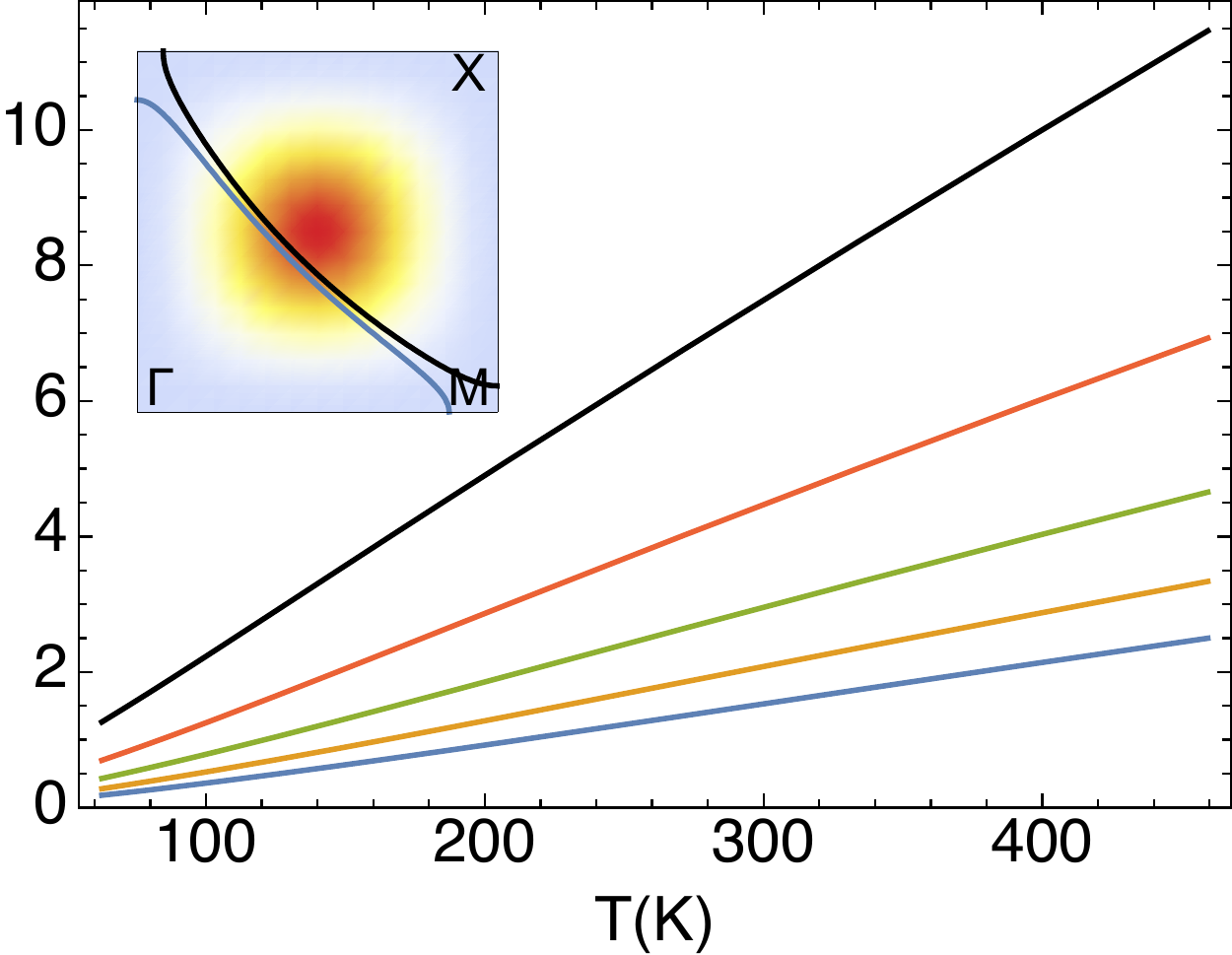}}
\caption{ 
  \label{DCrhovaryd} 
  Electrical  and Raman resistivities from \disp{pseudo-1}  at $t'=-0.2$ with varying hole doping $\delta$, as marked. The T dependence of electrical  resistivity and  the  $A_{1g}$  resistivity are concave-down   at small $\delta$, while the $B_{1g}$ and  $B_{2g}$ resistivity are flat or concave-up. (Inset) The displayed  Fermi surfaces  at $\delta=$  $.12$, $.24$  locate the maxima of $\Upsilon(k,\omega)$. The  relevant squared vertices from \disp{vertices} are shown as a heat map. The hot spots are movable by varying $t'$ and $t$.
  } 
 \end{figure*}

 The ``pseudo-identity'',  a statement of universality relating electrical transport and the dc limit of Raman intensities    noted by SS in \refdisp{raman1,raman2} is arrived at, if 
 we   assume  that $\rho_\alpha$ has a similar $T$ dependence for all $\alpha$:
 $
{\cal I}_\alpha(0,0)\sim  C_\alpha \; \frac{T}{ \rho_{xx}(T)} \label{pseudo-identity}
 $
 $C_\alpha$ is an $\alpha$-dependent constant.     Thus a $\rho\sim T^\sigma$ behavior would give rise to a $T^{1-\sigma}$ behavior for the Raman intensity in all channels.  We see in \figdisp{DCrhovaryd} that this suggestion  is true for the $A_{1g}$ resistivity at hole dopings, but it needs to be adjusted to the different $k$-dependent  filters that make the $B_{1g}$ and $B_{2g}$ channels different from the others. Thus the we limit the  universality of the  pseudo-identity in this work, and we quantify the effects of the bare vertices in the relationship between the members of the quartet of susceptibilities.  
 
  \begin{figure}[!]
\subfigure{\includegraphics[width=.7\columnwidth]{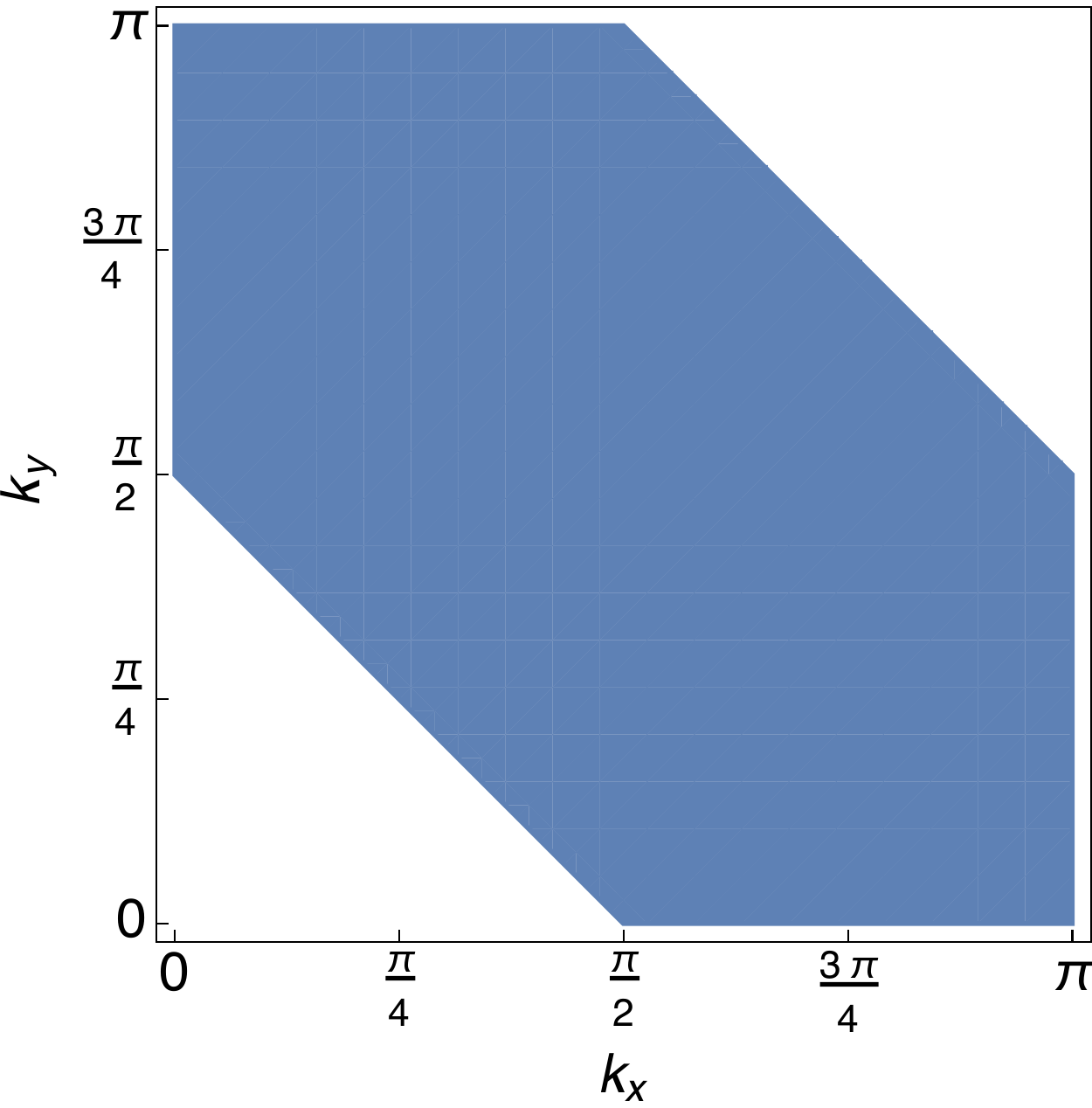}}
\caption{ 3
  \label{shaded} 
  Shaded region for estimating average scale of vertices. 
  }
 \end{figure}
 
  \begin{figure*}[!]
\subfigure[{Electrical resistivity $\bar{\rho}_{xx}$}]{\includegraphics[width=.7\columnwidth]{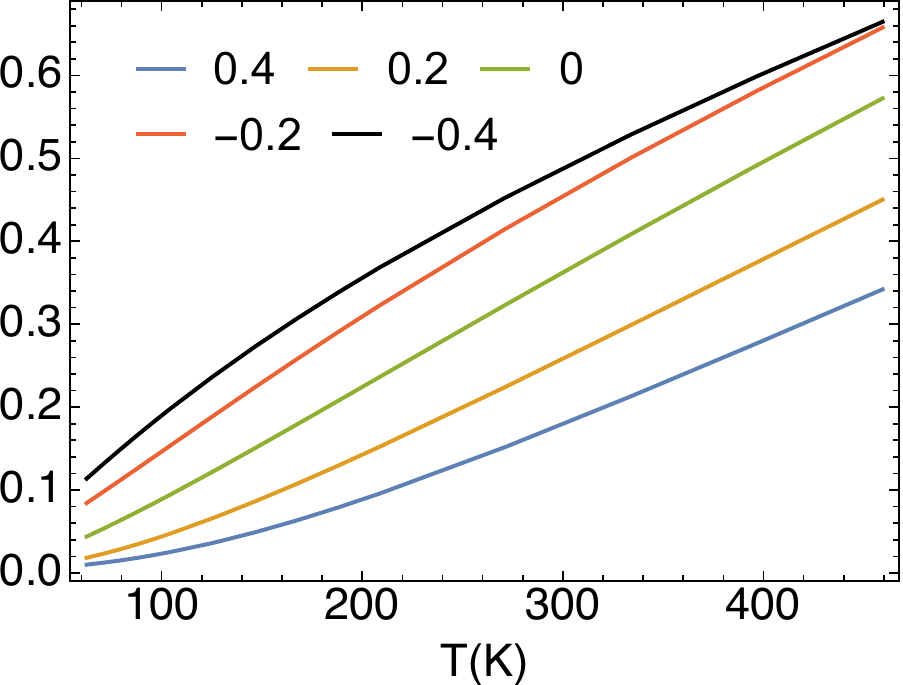}}
\subfigure[{$A_{1g}$ Raman: resistivity $\bar{\rho}_{A_{1g}}$}]{\includegraphics[width=.7\columnwidth]{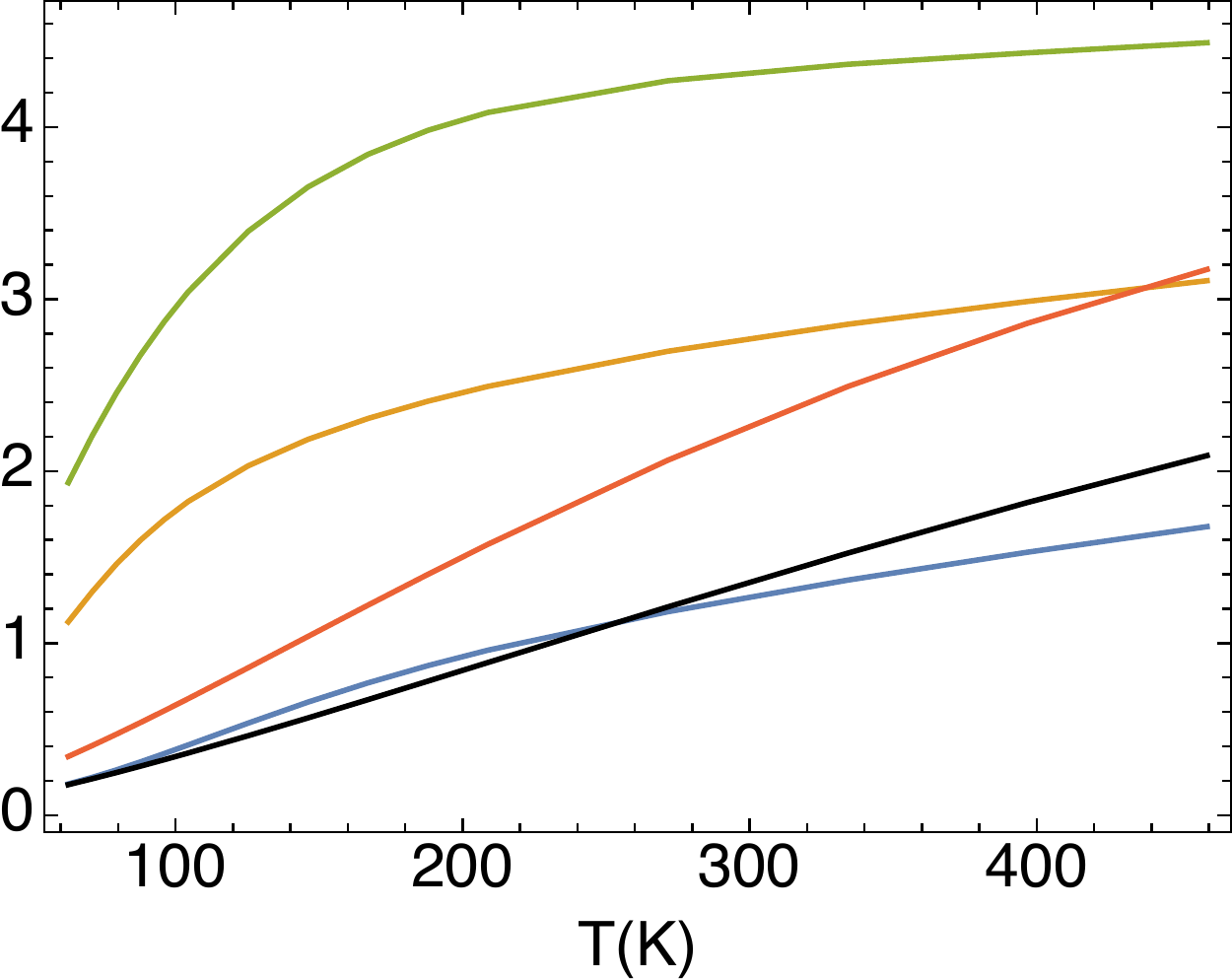}}
\subfigure[{$B_{1g}$ Raman: resistivity $\bar{\rho}_{B_{1g}}$}]{\includegraphics[width=.7\columnwidth]{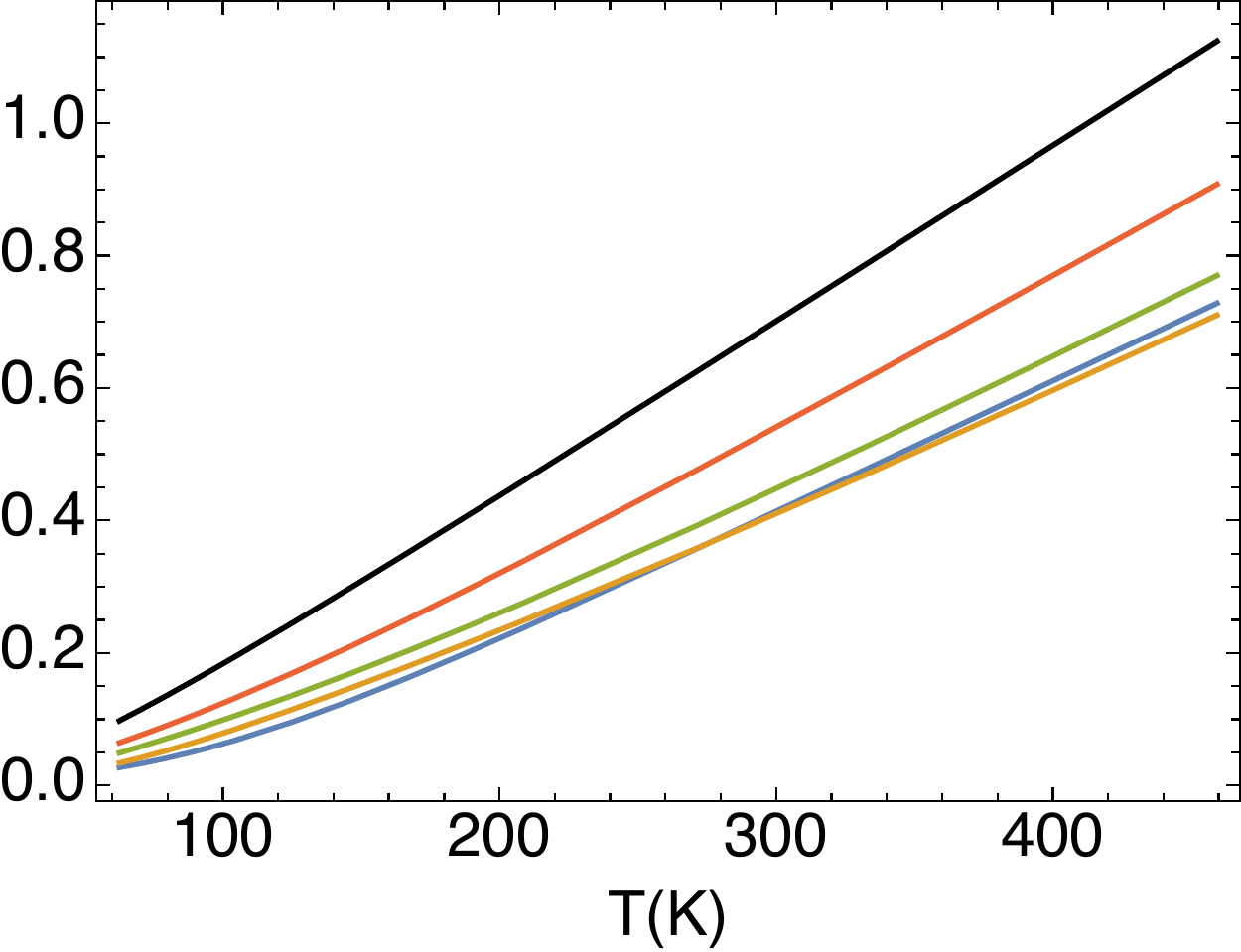}}
\subfigure[{$B_{2g}$ Raman: resistivity $\bar{\rho}_{B_{2g}}$}]{\includegraphics[width=.7\columnwidth]{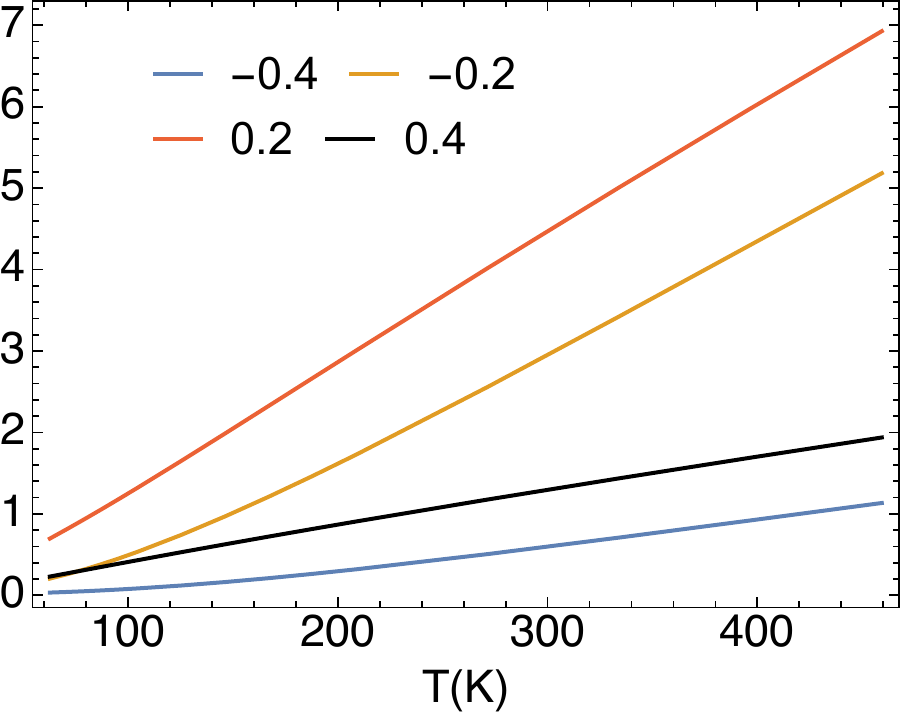}}
\caption{ 
  \label{DCrhovarytp} Dimensionless $\bar{\rho}_{xx}$, taken from \refdisp{SP}, $\bar{\rho}_{A_{1g}}$, $\bar{\rho}_{B_{1g}}$ and $\bar{\rho}_{B_{2g}}$  at $\delta=0.15$ with varying second neighbor hopping $t'$, as marked (same legend for all subfigures). \refdisp{Hackl} displays data  corresponding to the $B_{1g}$ geometry.}
 \end{figure*}

 Proceeding further using the bubble scheme we get {the imaginary part of} the dimensionless susceptibility 
$\bar{ \chi}''_\alpha(0,\omega)\equiv  \frac{ c_0 h}{N_s}  { \chi}''_\alpha(0,\omega)$  as
\beq
 \bar{ \chi}''_\alpha(0,\omega) =\omega \langle\Upsilon(k,\omega) {\cal J}^2_{\alpha}(k) \rangle_k, \label{DC}
\eeq
where $c_0 \sim 6.64{\mbox{\normalfont\AA}}$
is  a typical interlayer separation\cite{SP}.  The angular average is  $\langle A \rangle_k \equiv  \frac{1}{N_s}\sum_{k} A(k) $ and the momentum resolved relaxation scale is 
 \beq
\Upsilon(k,\omega) = \frac{4 \pi^2}{\omega}\int_{-\infty}^\infty dy \,  \rho_G(k,y)\rho_G(k,y+\omega)  [f(y)-f(\omega+y)]. \nn \label{opconductivity}
\eeq
Here $\rho_G(k,\omega)$ is the electron spectral function.  With $\rho_{1,\alpha}\equiv \frac{c_0 h}{\zeta_\alpha}$, the corresponding 
dimensionless 
conductivity  $\bar{\sigma}_\alpha(\omega)\equiv\rho_{1,\alpha} \times \sigma_\alpha(\omega)$
is given by
\beq
 \bar{\sigma}_\alpha(\omega)=\langle\Upsilon(k,\omega) {\cal J}^2_{\alpha}(k) \rangle_k. \label{optical}
\eeq
From Eqs~(\ref{DC}) and (\ref{optical}), we can see $\bar{ \chi}''_\alpha(0,\omega)=\omega*\bar{\sigma}_\alpha(\omega)$

\section{ Parameter region}
We explore how the variation of second neighbor hopping $t'$, doping $\delta$ and temperature $T$ affects the quartet of conductivities and susceptibilities in the normal state. We focus on optimal doping or slightly overdoped cases from electron-doped (positive $t'$) to hole-doped (negative $t'$) systems. Our temperature region starts from 63K to a few hundred Kelvin.

\section{\bf DC  limit and electrical resistivity results:} \label{DCT}

Using the spectral function from the second order ECFL theory, we calculate the dimensionless DC ($\omega\rightarrow0$) electrical and Raman conductivities $\bar{\sigma}_{\alpha}$ from \disp{optical}. The corresponding dimensionless resistivities are  
\beq
\bar{\rho}_\alpha=\frac{1}{\bar{\sigma}_{a}}=\frac{1}{\langle\Upsilon(k,0) {\cal J}^2_{\alpha}(k) \rangle_k},\label{resistivity}
\eeq
The  electrical resistivity in physical units is given by $\rho_{xx}= \bar{\rho}_{xx} \times \rho_{1,xx}$, with $\rho_{1, xx}= c_0 \frac{h}{e^2}\; \sim 1.71\, m \, \Omega$cm\cite{SP}.

We calculate typical quantities for the three Raman geometries and the electrical conductivity from \disp{vertices} as a set of quartets below. The comparison of the figures in each set is of interest, since the different functions in the bare vertices pick out different parts of the $k$-space. 
In this paper  $t=1$ serves as the energy unit;  for the systems in mind we estimate \cite{SP}  $t\sim .45$ eV.

\begin{figure*}[ht]
\subfigure[{Optical conductivity $\bar{\sigma}_{xx}$ and susceptibility $\bar{\chi}''_{xx}$}]{\includegraphics[width=.7\columnwidth]{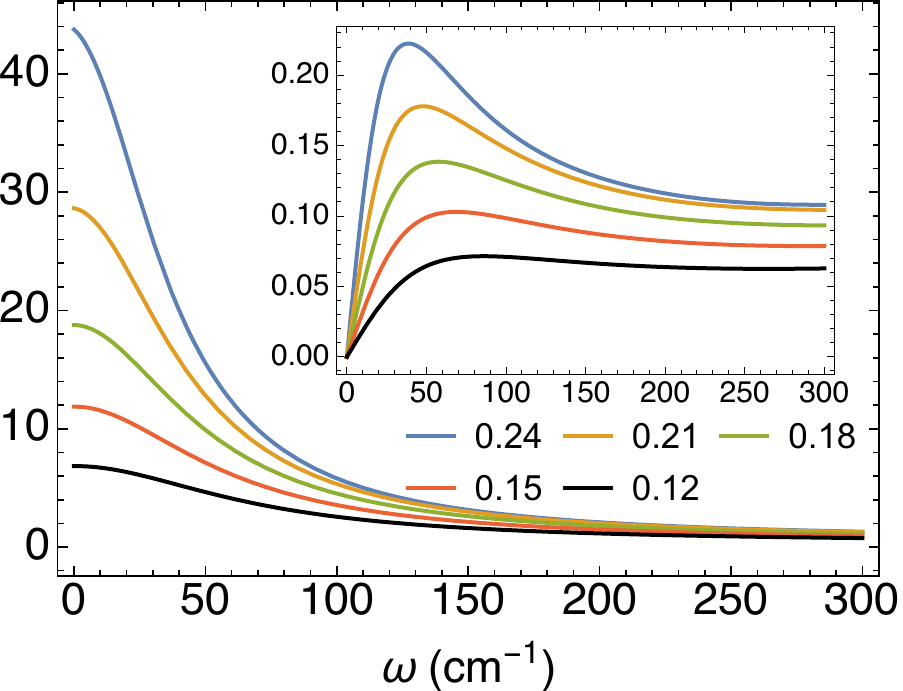}}
\subfigure[{$A_{1g}$ Raman: $\bar{\sigma}_{A_{1g}}$ and $\bar{\chi}''_{A_{1g}}$ (inset)}]{\includegraphics[width=.7\columnwidth]{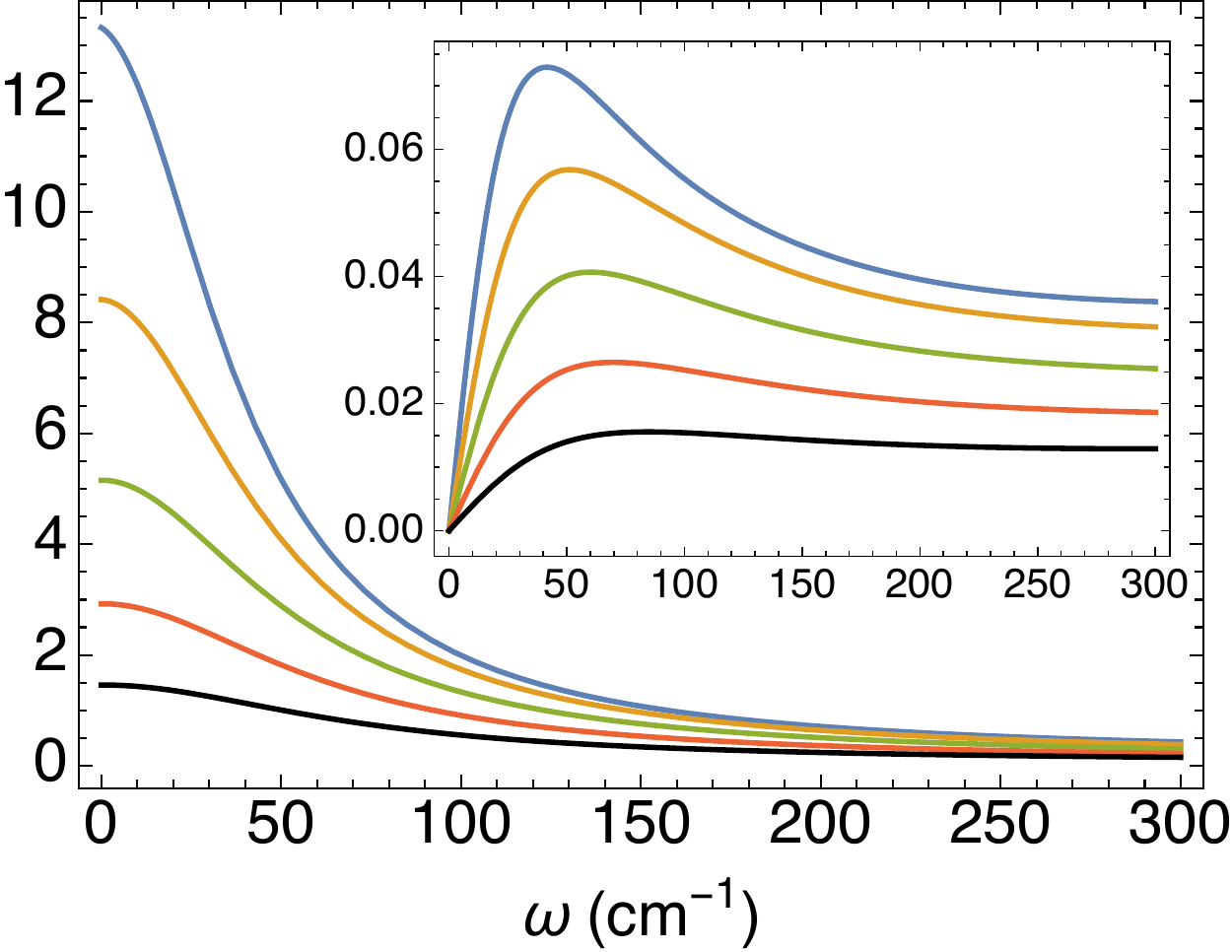}}
\subfigure[{$B_{1g}$ Raman: $\bar{\sigma}_{B_{1g}}$ and $\bar{\chi}''_{B_{1g}}$ (inset)}]{\includegraphics[width=.7\columnwidth]{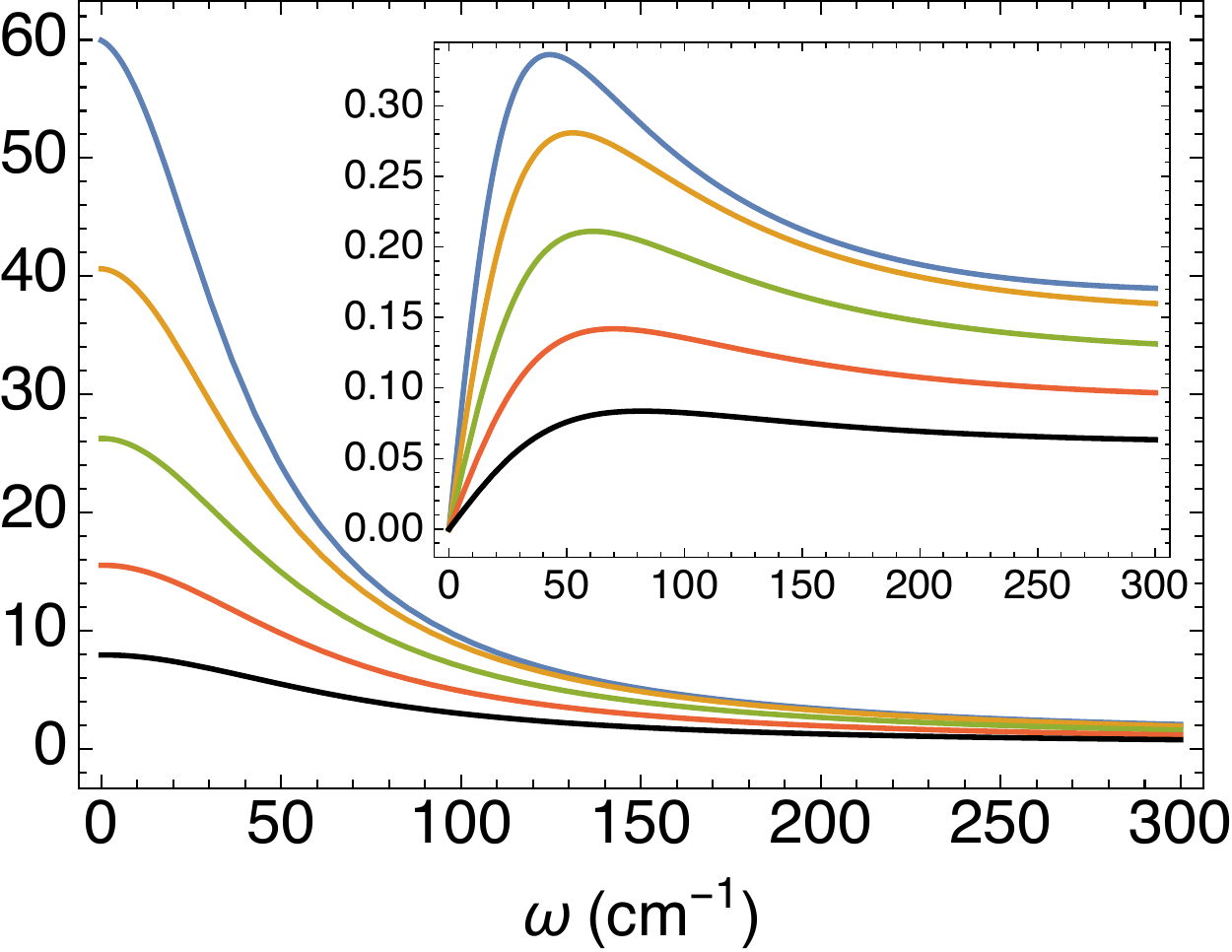}}
\subfigure[{$B_{2g}$ Raman: $\bar{\sigma}_{B_{2g}}$ and $\bar{\chi}''_{B_{2g}}$ (inset)}]{\includegraphics[width=.7\columnwidth]{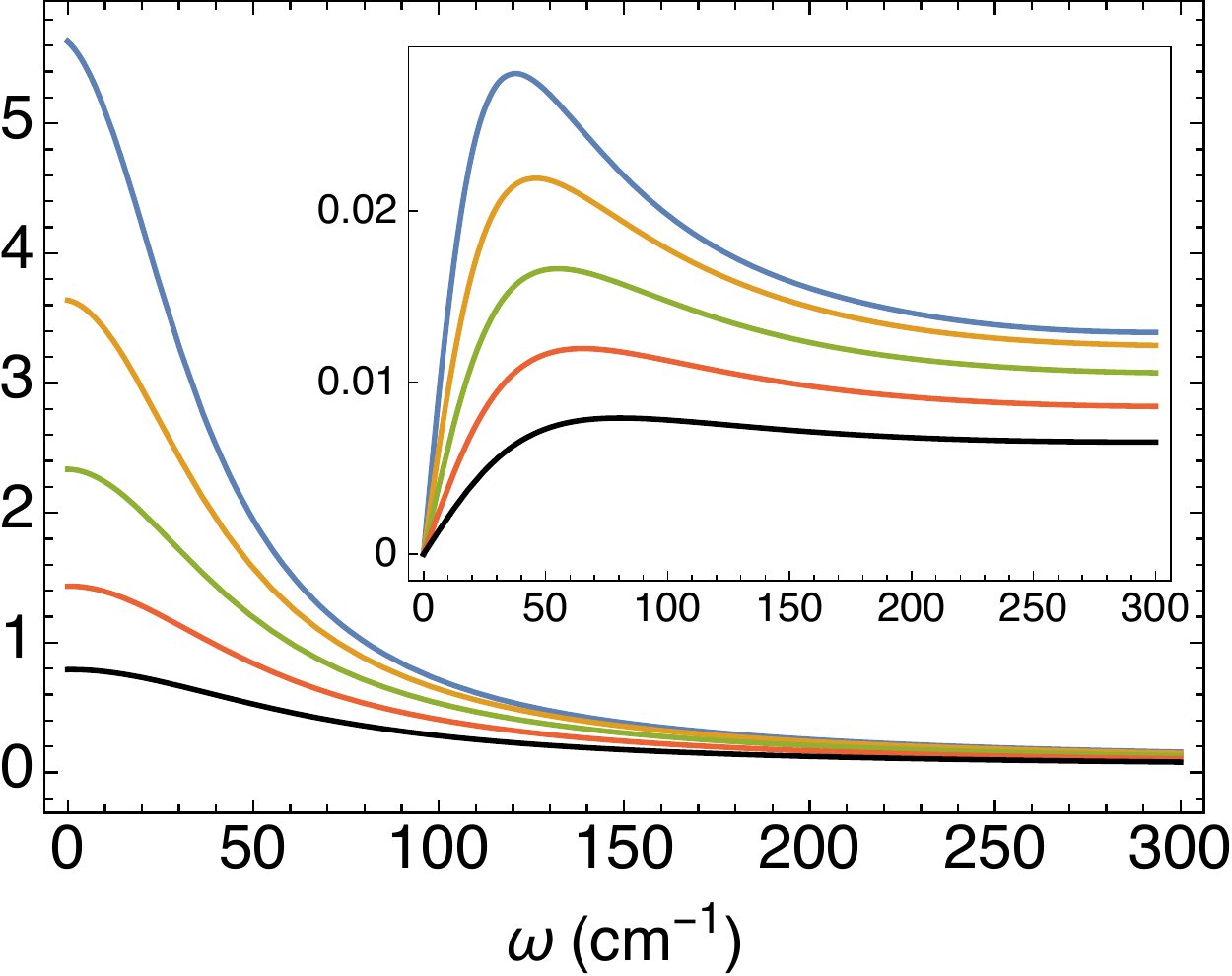}}
 \caption{ 
  \label{opsigmatpn2varyd} Dynamical conductivities $\bar{\sigma}_\alpha$ and susceptibilities  $\bar{\chi}_\alpha''$ (inset), for the hole doped case  $t'=-0.2$, $T=63$K  at different $\delta$, as marked. In the experiments in \refdisp{Sugai-2} Fig.~(1), the same  quartet of results is shown for LSCO.
   At the highest energy of over $1000$K, as in the data,   the  susceptibility shows no sign of dropping off. }
 \end{figure*}

\begin{figure*}[ht]
\subfigure[{Optical conductivity $\bar{\sigma}_{xx}$ and susceptibility $\bar{\chi}''_{xx}$ (inset)}]{\includegraphics[width=.7\columnwidth]{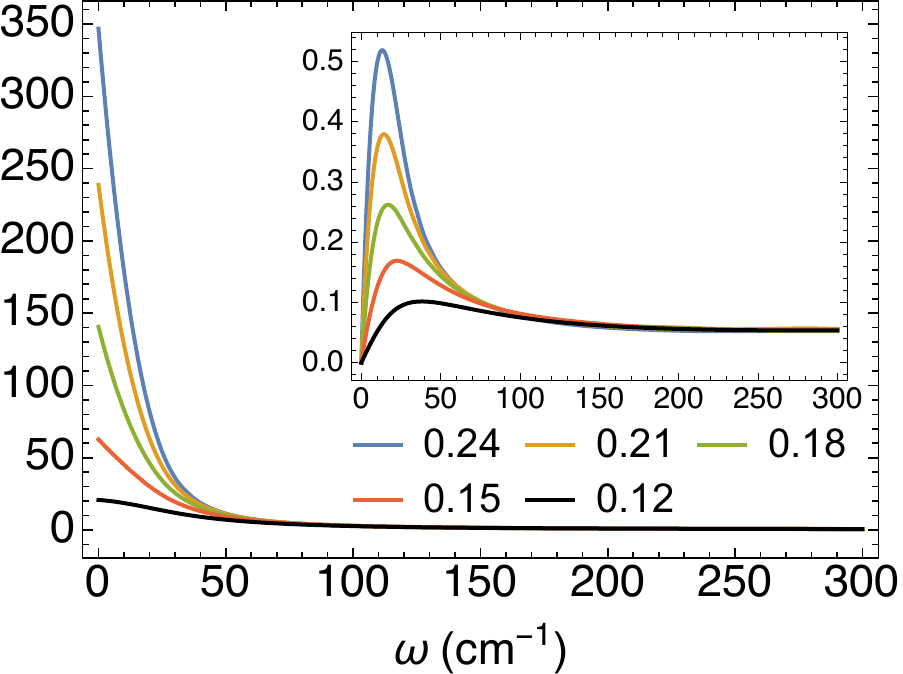}}
\subfigure[{$A_{1g}$ Raman: $\bar{\sigma}_{A_{1g}}$ and $\bar{\chi}''_{A_{1g}}$ (inset)}]{\includegraphics[width=.7\columnwidth]{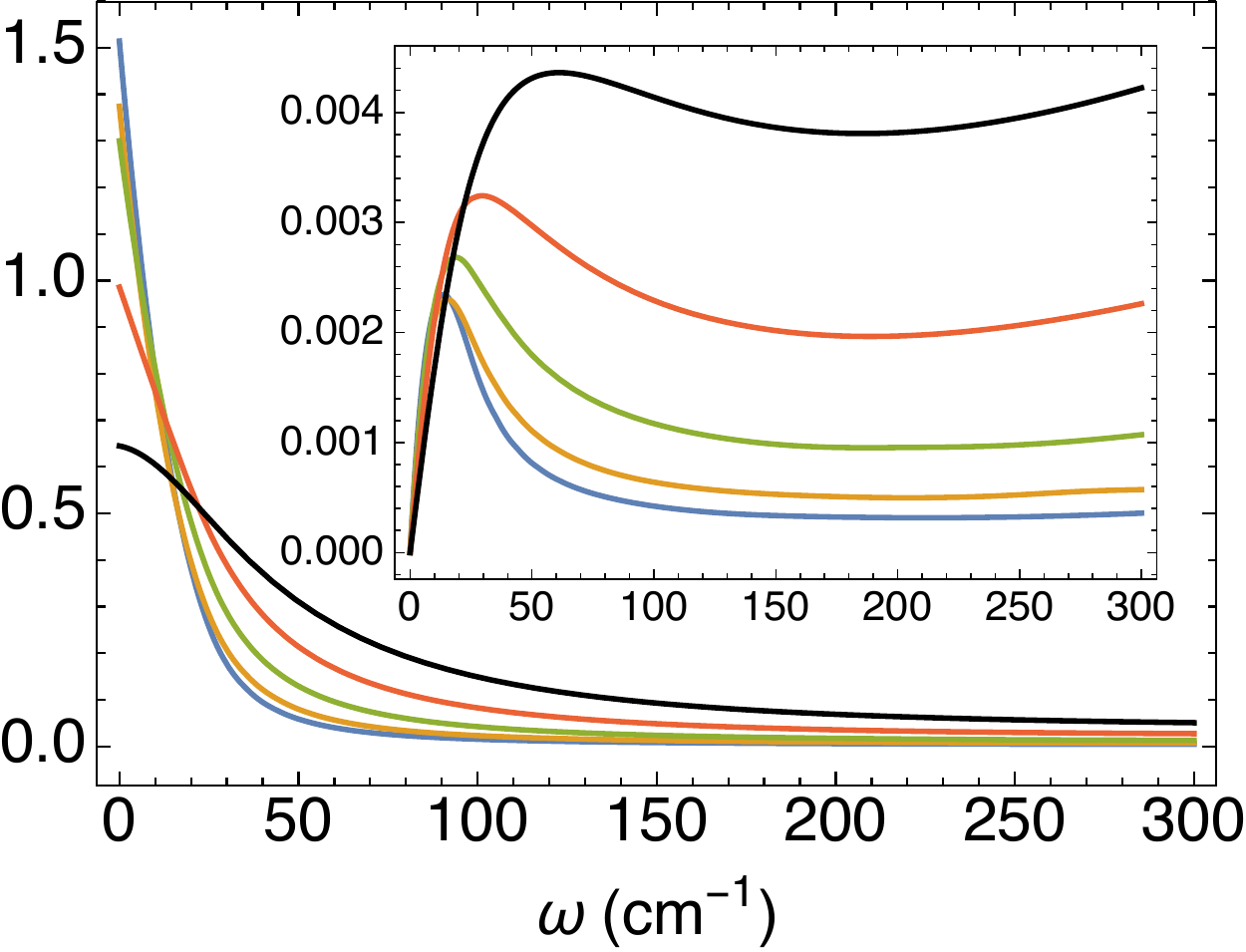}}
\subfigure[{$B_{1g}$ Raman: $\bar{\sigma}_{B_{1g}}$ and $\bar{\chi}''_{B_{1g}}$ (inset)}]{\includegraphics[width=.7\columnwidth]{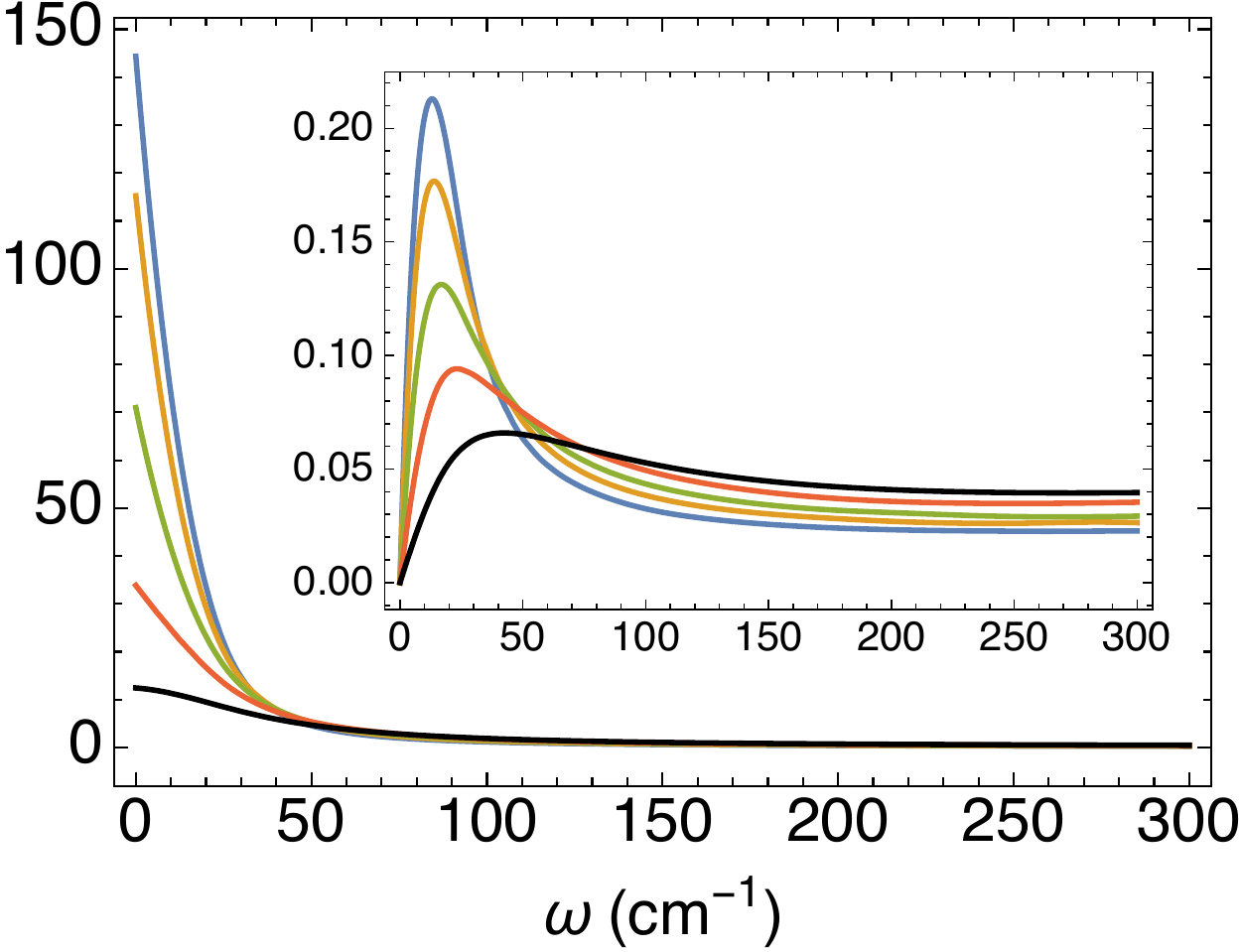}}
\subfigure[{$B_{2g}$ Raman: $\bar{\sigma}_{B_{2g}}$ and $\bar{\chi}''_{B_{2g}}$ (inset)}]{\includegraphics[width=.7\columnwidth]{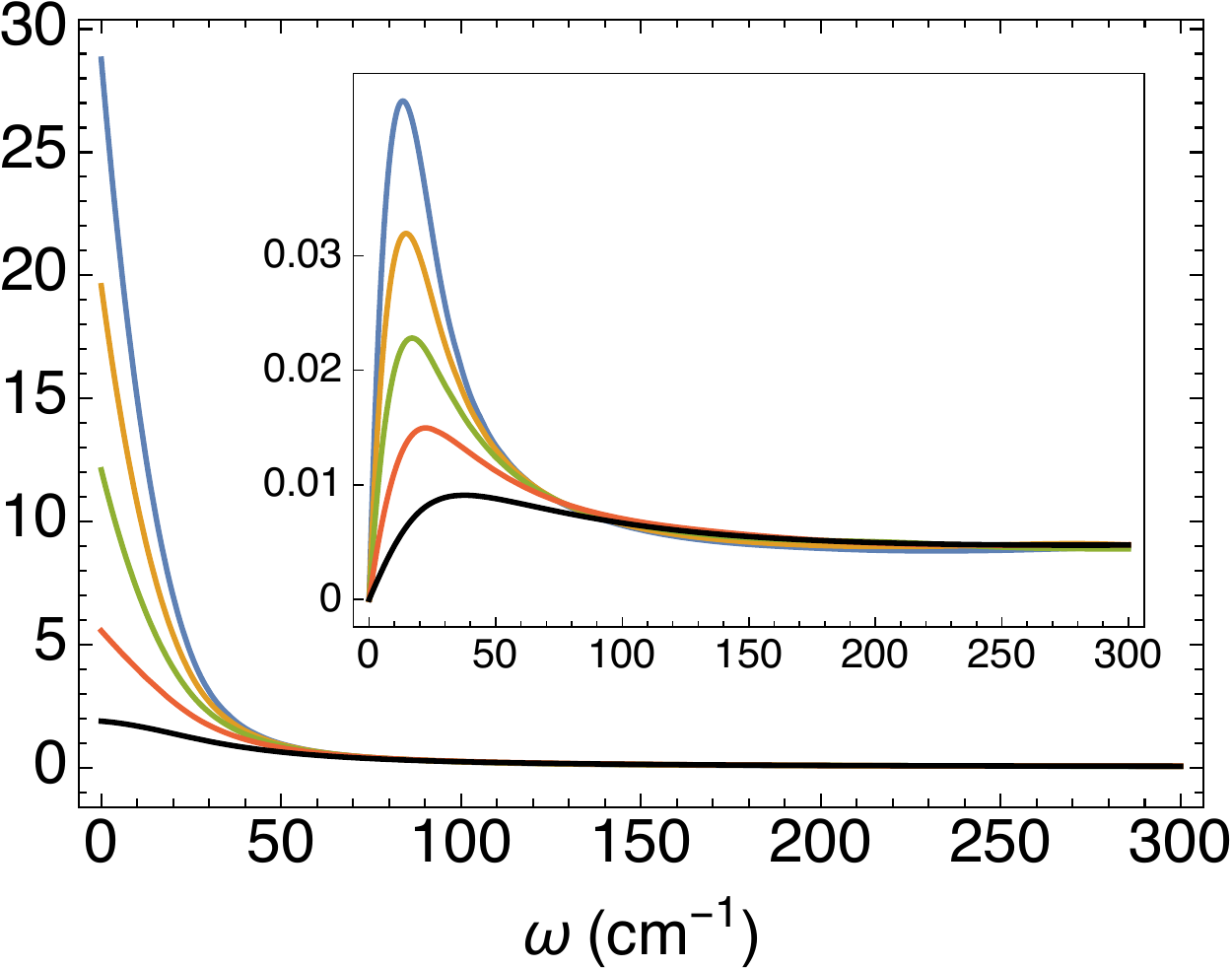}}
 \caption{ 
  \label{opsigmatpp2varyd} Dynamical conductivities $\bar{\sigma}_\alpha$ and susceptibilities  $\bar{\chi}_\alpha''$ (inset), for the electron doped case  $t'=0.2$, $T=63$K  at different $\delta$, as marked.}
 \end{figure*}

In \figdisp{DCrhovaryd}, we plot DC resistivity $\bar{\rho}_{xx}$ and Raman resistivities in the DC limit $\bar{\rho}_{A_{1g}}$, $\bar{\rho}_{B_{1g}}$, $\bar{\rho}_{B_{2g}}$ varying hole doping $\delta$ and fixing $t'=-0.2$. The four figures  have roughly similar doping dependence,  as suggested by  the pseudo-identity. They all decrease when the doping increases, although the curvature changes more in $\bar{\rho}_{xx}$ and $\bar{\rho}_{A_{1g}}$ than the other two cases. This can be understood from Eq. (\ref{DC}) since  they  arise from  the same kernel $\Upsilon(k,0)$ with different filters. The quasi-particle peak in $\rho_G$, contributing most to $\Upsilon(k,0)$, is located along the Fermi surface and gets broadened when warming up. The inset shows the corresponding squared vertex ${\cal J}^2_{\alpha}$ in the background and the Fermi surfaces at the lowest and highest dopings. The $B_{1g}$ vertex vanishes along the line $k_x=k_y$ while the $B_{2g}$ vertices vanish near  $\{\pi,0\}$ and $\{0, \pi\}$ points. In our calculation both $B_{1g}$ and $B_{2g}$ overlap well with the peak region of the spectral function, whereas $A_{1g}$ and the resistivity do not. This results in the difference between the T dependence of them and the other two in \figdisp{DCrhovaryd}. It would be of considerable interest to study this pattern of T dependences systematically in future Raman studies. 

Although all $\bar{\rho}_{\alpha}$ increase when reducing doping $\delta$ approaching the half-filling limit due to the suppression of quasi-particles, their magnitudes at high temperature vary considerably, as a result of different vertices filtering the contribution from $\Upsilon(k,0)$. We can understand this scale difference by evaluating the average of vertices over the shaded region in \figdisp{shaded}. The shaded region covers the Fermi surface for all chosen $\delta$ and $t'$, and therefore contains the most significant contribution to $\rho_\alpha$. 

At $t'=-0.2$, $\langle {\cal J}^2_{xx} \rangle_s\approx2.41$, $\langle {\cal J}^2_{A_{1g}} \rangle_s\approx0.56$, $\langle {\cal J}^2_{B_{1g}} \rangle_s\approx1.30$, $\langle {\cal J}^2_{B_{2g}} \rangle_s\approx0.20$, where $\langle  \rangle_s$ represents the $k$ average over the shaded region. They not only explain the relation $\bar{\rho}_{xx}<\bar{\rho}_{B_{1g}}<\bar{\rho}_{A_{1g}}<\bar{\rho}_{B_{2g}}$, but also capture the ratio among them rather closely at high enough T. The structure at low T is more subtle, and carries information about the magnitude of $t'$  that cannot be captured by the above high T argument.
 
Although all $\bar{\rho}_\alpha$ increase as $\delta$ decreases in general, their $t'$ dependence can be rather different, as shown in \figdisp{DCrhovarytp}. $\bar{\rho}_{xx}$ and $\bar{\rho}_{B_{1g}} $ decreases monotonically in general as $t'$ increases from hole-doped (negative) to electron-doped (positive), while $\bar{\rho}_{A_{1g}}$ and $\bar{\rho}_{B_{2g}}$ decrease only as $|t'|$ increases and their monotonicity with respect to $t'$ changes upon sign change of $t'$. Another interesting observation is that $\bar{\rho}_\alpha(t'=-0.2)>\bar{\rho}_\alpha(t'=0.2)$ and $\bar{\rho}_\alpha(t'=-0.4)>\bar{\rho}_\alpha(t'=0.4)$ are true for $\alpha=\mbox{xx}$, $B_{1g}$ and $B_{2g}$, but for the $A_{1g}$ case, $\bar{\rho}_\alpha(t'=-0.2)<\bar{\rho}_\alpha(t'=0.2)$ in general and $\bar{\rho}_\alpha(t'=-0.4)\approx\bar{\rho}_\alpha(t'=0.4)$.

In \disp{resistivity}, the resistivities depend on $t'$ through $\Upsilon(k,0)$ and ${\cal J}^2_{\alpha}$. To estimate their $t'$ dependence, we can look at their average over the shaded region $\langle \Upsilon(k,0) \rangle_s$ and $\langle {\cal J}^2_{\alpha} \rangle_s$. While $\langle \Upsilon(k,0) \rangle_s$ rises monotonically as $t'$ increases, $\langle {\cal J}^2_{\alpha} \rangle_s$ ($\alpha=\mbox{xx}, A_{1g}, B_{2g}$) is a quadratic function of $t'$ which behaves differently at positive and negative $t'$, as shown in \disp{vertices}.

In the simplest $B_{1g}$ case, ${\cal J}^2_{B_{1g}}$ is independent of $t'$. Then $t'$ only affects $\bar{\rho}_{B_{1g}}$ through $\Upsilon(k,0)$ and therefore $\bar{\rho}_{B_{1g}}$ increases almost monotonically as $t'$ decreases (the crossing between $t=0.2$ and $t=0.4$ is due to the fact that the change on Fermi surface geometry leads to different filtering result when coupling to ${\cal J}^2_{B_{1g}}$). In the charge current case, the $t'$ dependence of $\Upsilon(k,0)$ still dominates since $\bar{\rho}_{xx}$ behaves similar to $\bar{\rho}_{B_{1g}}$ and the contribution from ${\cal J}^2_{xx}$ mostly modifies the curvature without affecting the relative scale.

 \begin{figure*}[ht]
\subfigure[{Optical conductivity $\bar{\sigma}_{xx}$ and susceptibility $\bar{\chi}''_{xx}$}]{\includegraphics[width=.7\columnwidth]{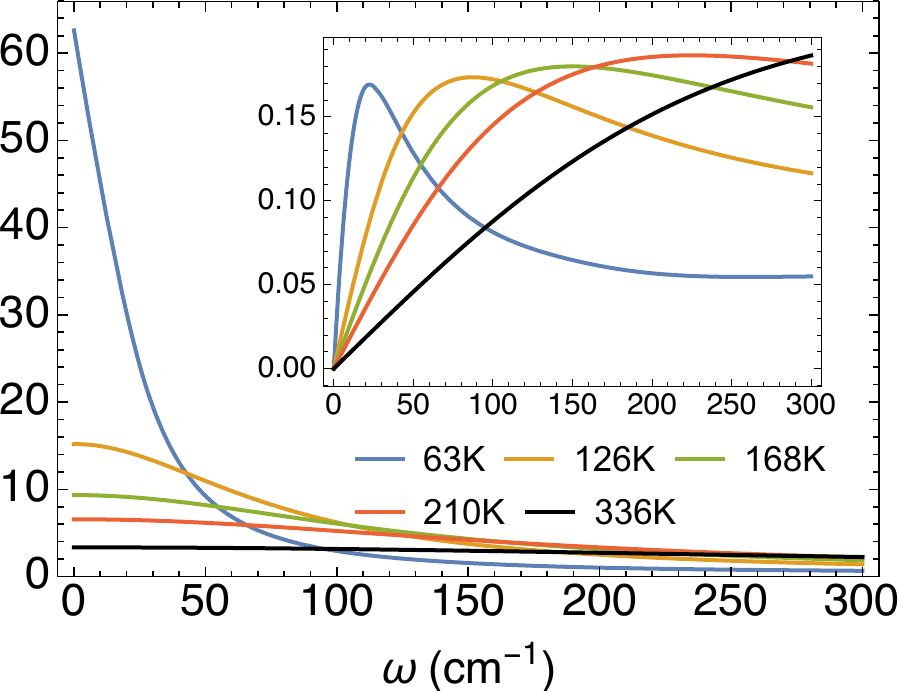}}
\subfigure[{$A_{1g}$ Raman: $\bar{\sigma}_{A_{1g}}$ and $\bar{\chi}''_{A_{1g}}$ (inset)}]{\includegraphics[width=.7\columnwidth]{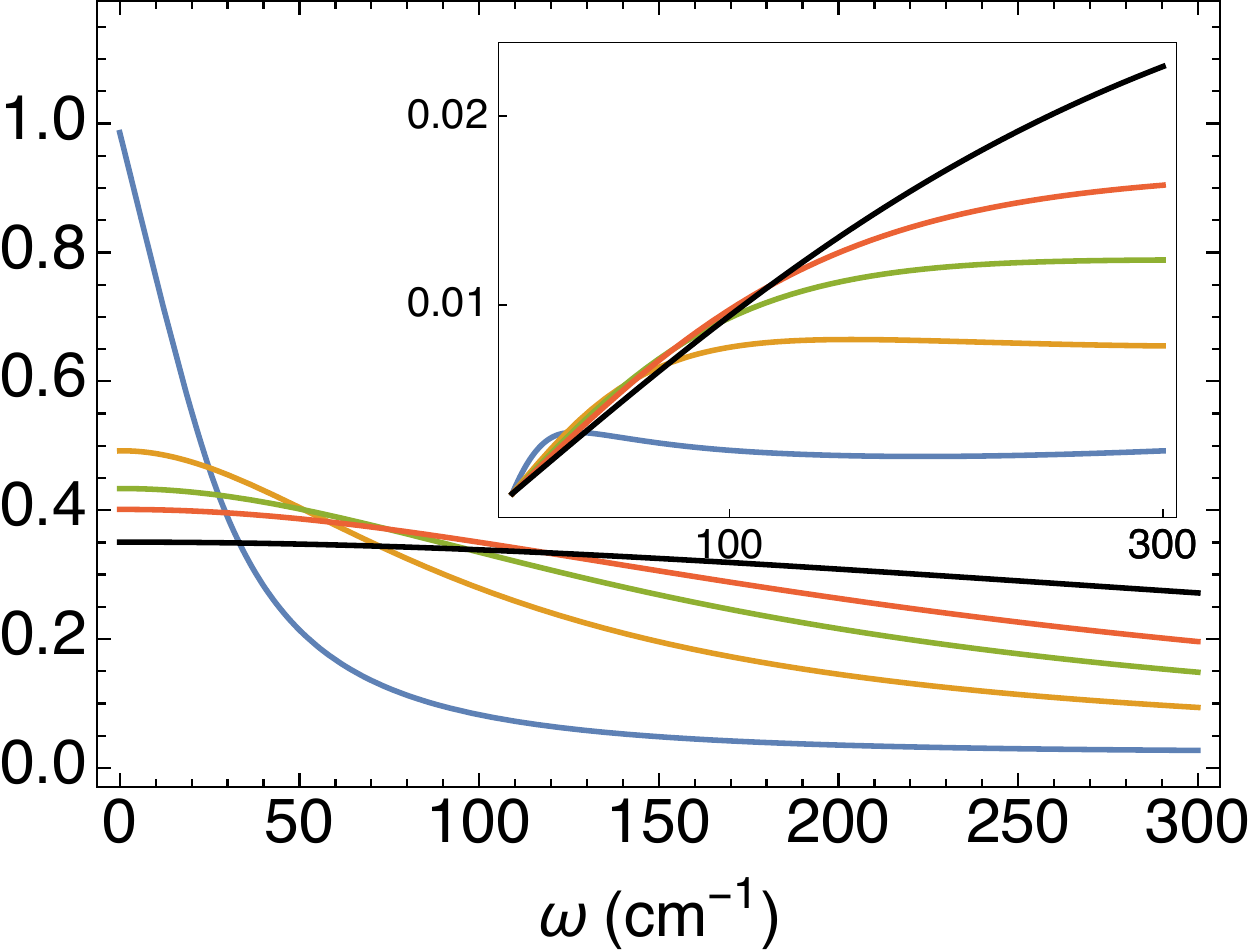}}
\subfigure[{$B_{1g}$ Raman: $\bar{\sigma}_{B_{1g}}$ and $\bar{\chi}''_{B_{1g}}$ (inset)}]{\includegraphics[width=.7\columnwidth]{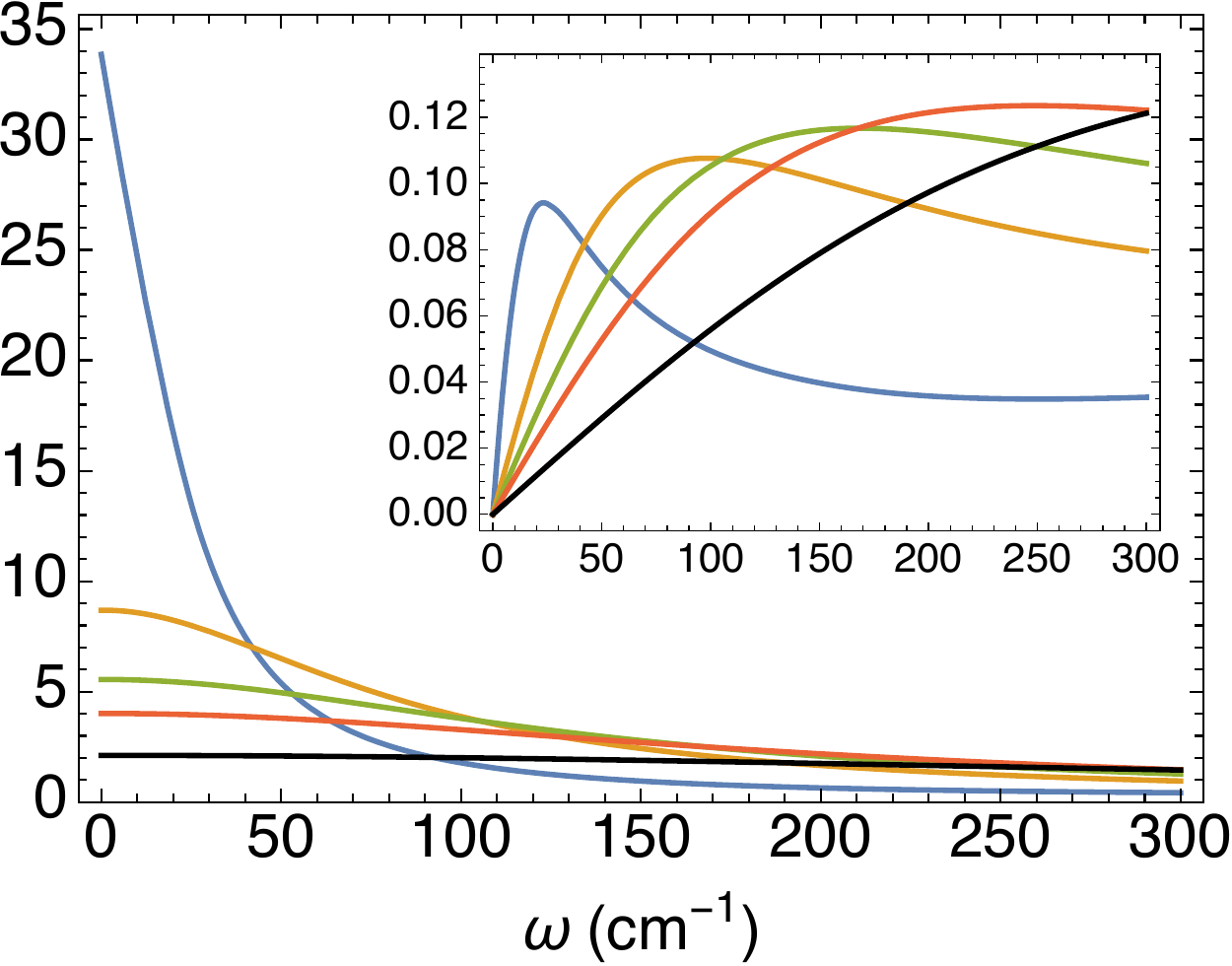}}
\subfigure[{$B_{2g}$ Raman: $\bar{\sigma}_{B_{2g}}$ and $\bar{\chi}''_{B_{2g}}$ (inset)}]{\includegraphics[width=.7\columnwidth]{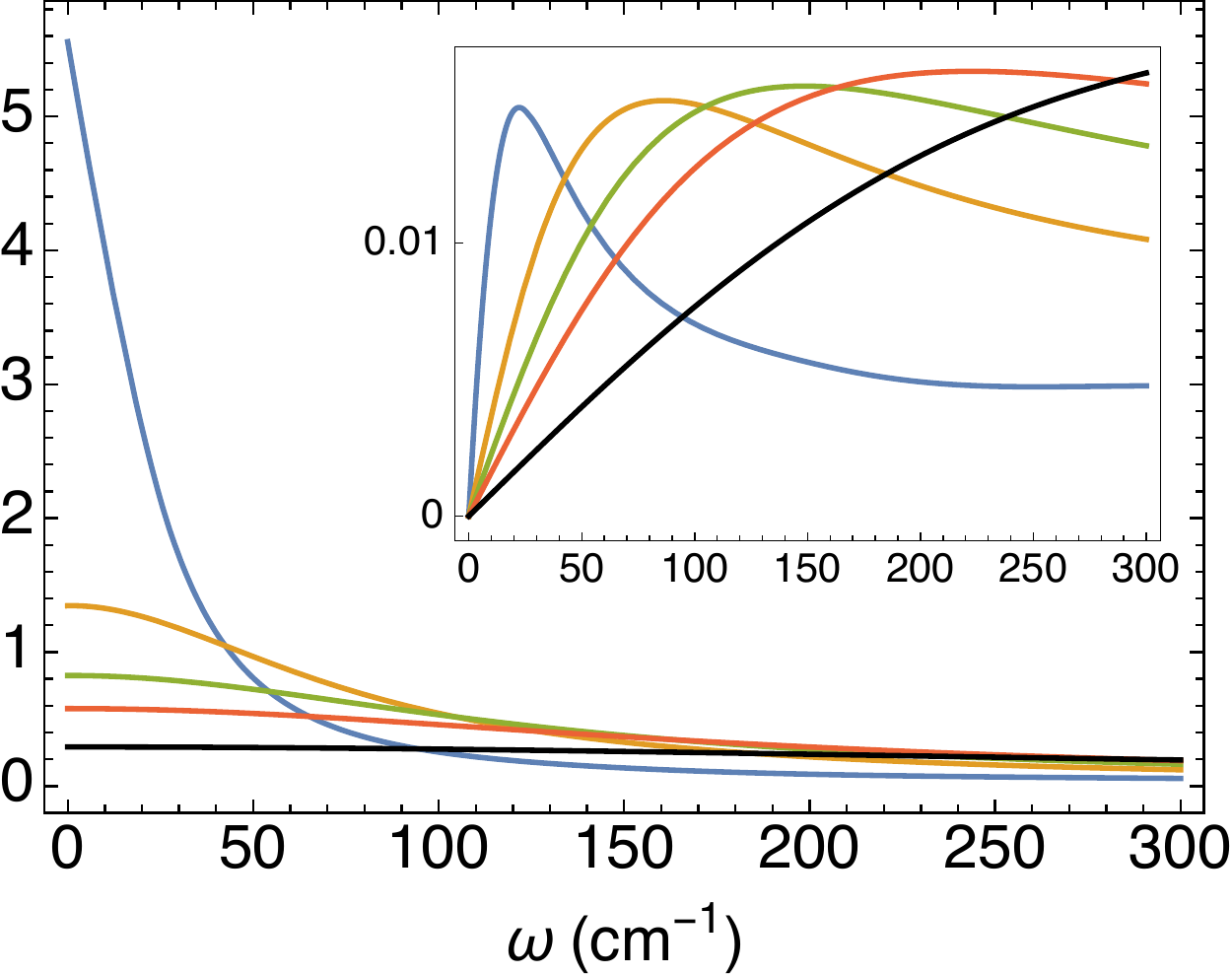}}
 \caption{ 
  \label{tpp2opsigmavaryT} Dynamical conductivities and (inset) susceptibilities for the electron doped case with  $t'=0.2$, $\delta=0.15$, for  various  $T$'s as marked.   Subfigure(d)  with $B_{2g}$ symmetry is comparable to the high resolution experimental result in  Fig.~(2) of   \refdisp{Koitzsch} at comparable set of $T$'s. The theoretical curve reproduces well the quasi-elastic peaks, and  their  T evolution.   }
 \end{figure*}
 
  \begin{figure*}[!]
\subfigure[{Optical conductivity $\bar{\sigma}_{xx}$ and susceptibility $\bar{\chi}''_{xx}$}]{\includegraphics[width=.7\columnwidth]{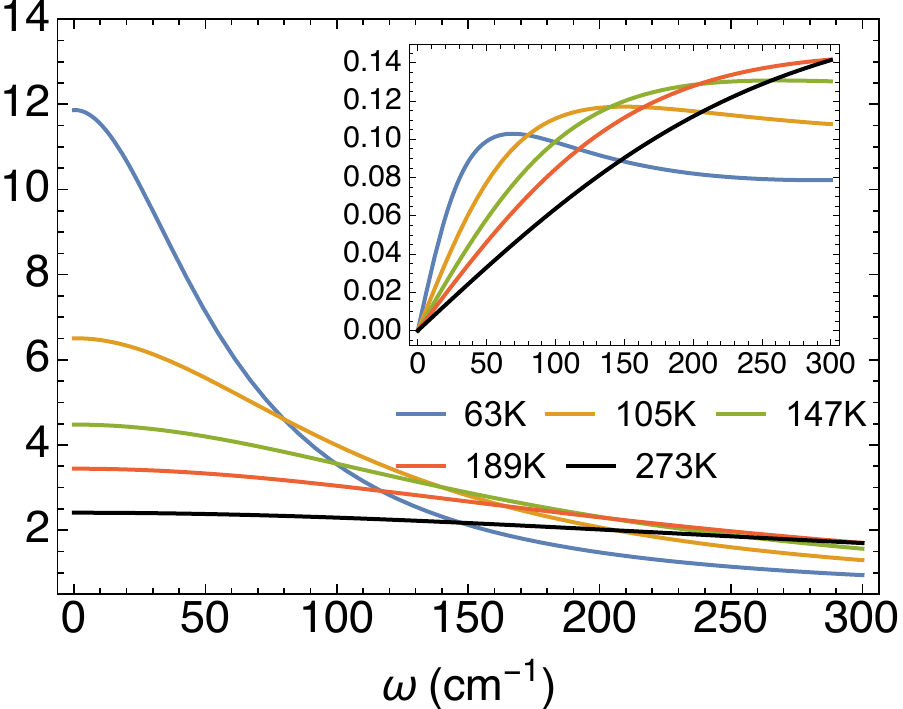}}
\subfigure[{$A_{1g}$ Raman: $\bar{\sigma}_{A_{1g}}$ and $\bar{\chi}''_{A_{1g}}$ (inset)}]{\includegraphics[width=.7\columnwidth]{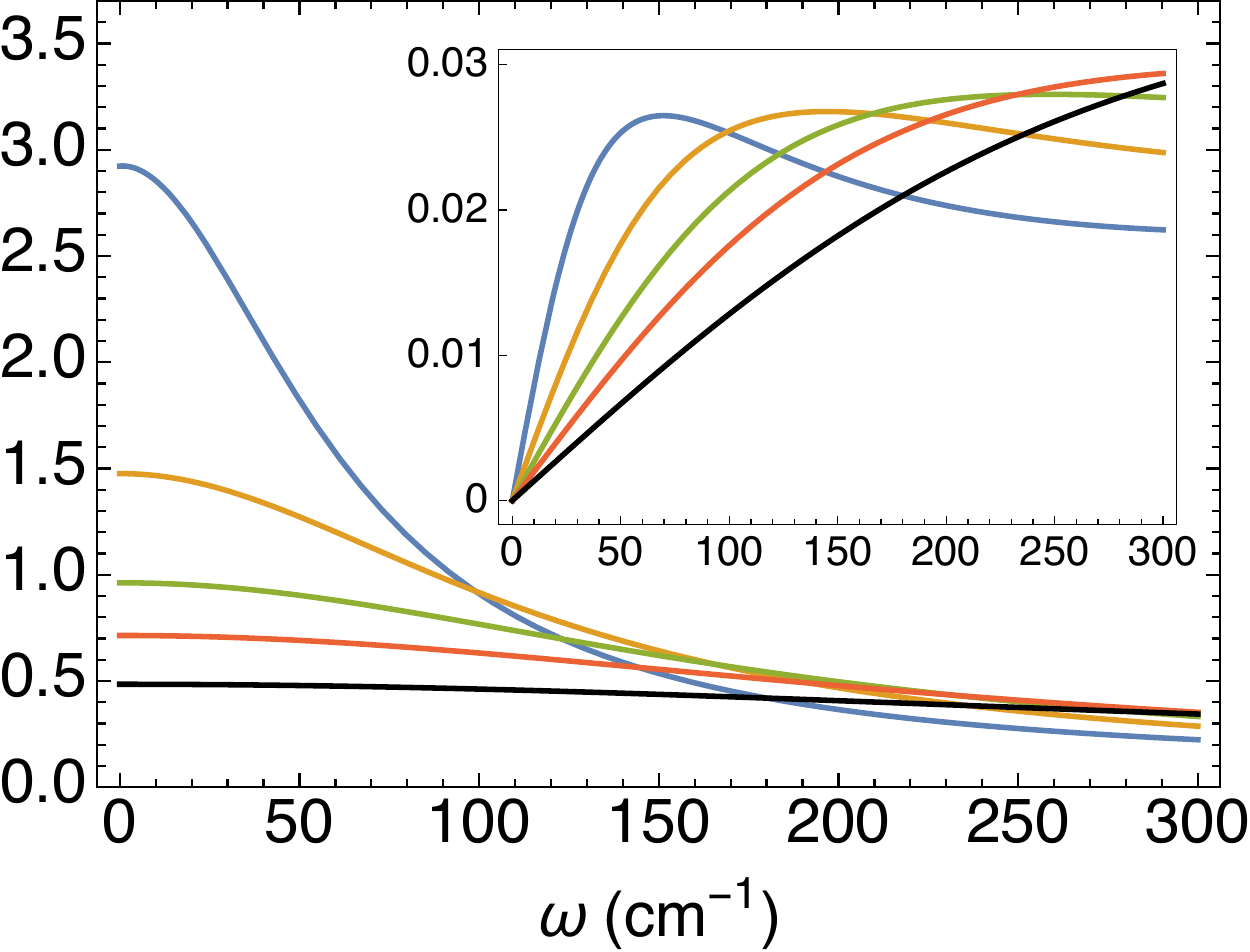}}
\subfigure[{$B_{1g}$ Raman: $\bar{\sigma}_{B_{1g}}$ and $\bar{\chi}''_{B_{1g}}$ (inset)}]{\includegraphics[width=.7\columnwidth]{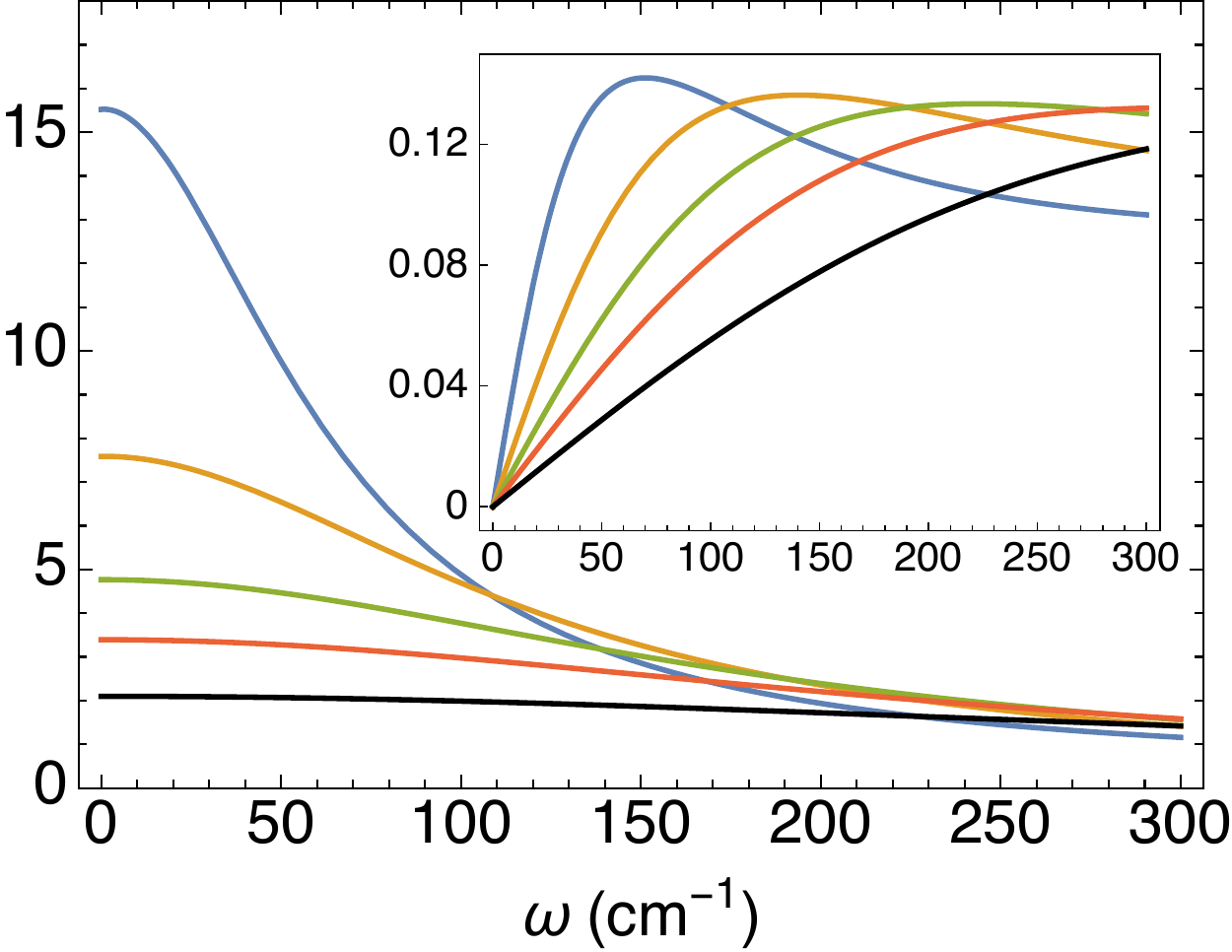}}
\subfigure[{$B_{2g}$ Raman: $\bar{\sigma}_{B_{2g}}$ and $\bar{\chi}''_{B_{2g}}$ (inset)}]{\includegraphics[width=.7\columnwidth]{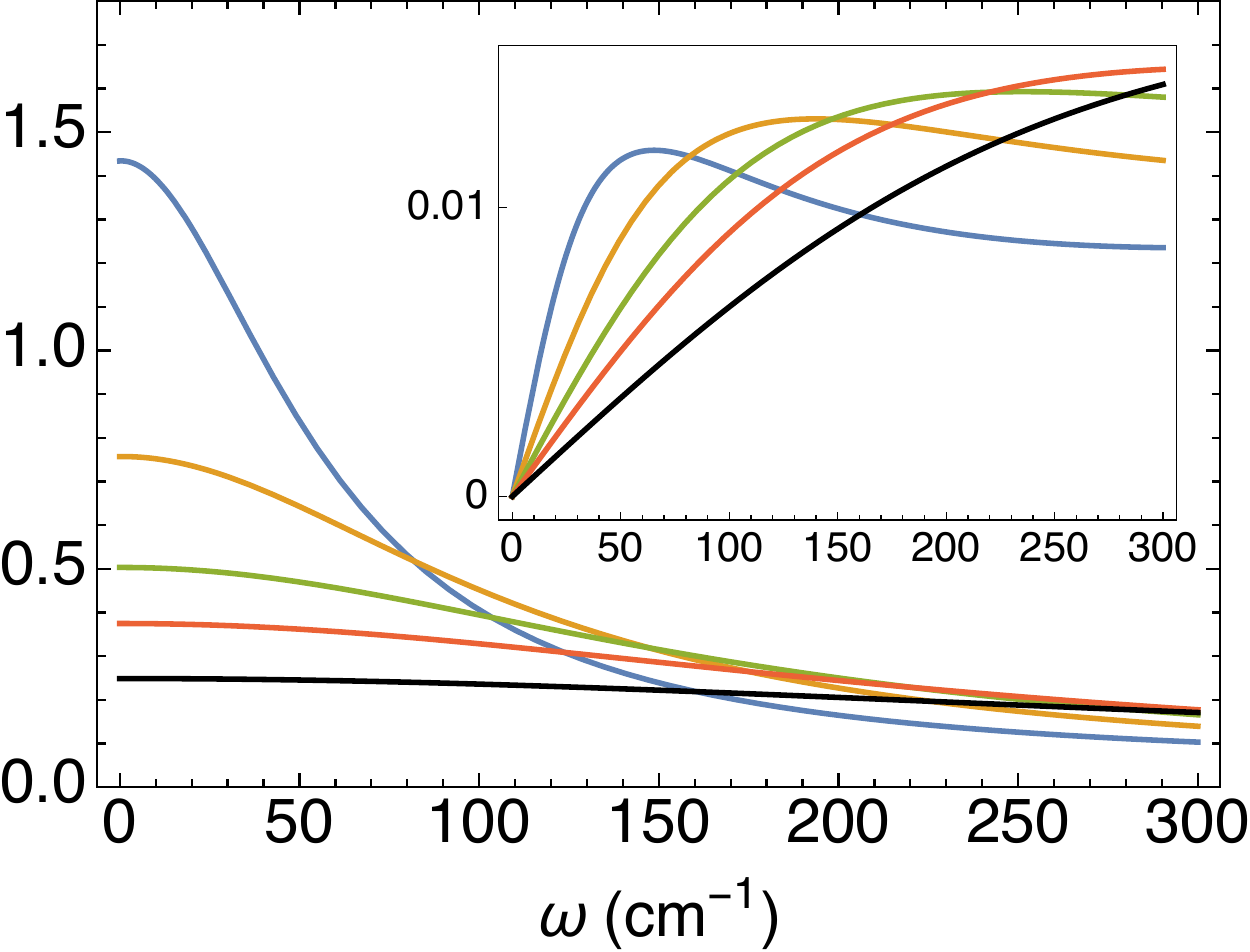}}
 \caption{ 
  \label{opsigmavaryT} Optical conductivity $\bar{\sigma}_{xx}(\omega)$ and the Raman conductivities $\bar{\sigma}_{A_{1g}}(\omega)$, $\bar{\sigma}_{B_{1g}}(\omega)$, $\bar{\sigma}_{B_{2g}}(\omega)$ at $t'=-0.2$, $\delta=0.15$ and varying $T$, as marked (same legend for all subfigures). The corresponding dimensionless susceptibility is plotted in the inset with the same x-axis. \refdisp{Sugai-2}, \refdisp{Sugai-3} and \refdisp{Koitzsch} show data that corresponds to these variables. }
 \end{figure*}

The different behaviors in the other two cases indicate the quadratic $t'$ dependence in ${\cal J}^2_{\alpha}$ ($\alpha=A_{1g}, B_{2g}$) becomes dominant. In the simpler $B_{2g}$ case, ${\cal J}^2_{B_{2g}}\propto t'^2$ provides the dominant $t'$ dependence in $\bar{\rho}_{B_{2g}}$, explaining $\bar{\sigma}_{B_{2g}}(t'=0)=0$ and $\bar{\rho}_{B_{2g}}(|t'|=0.2)>\bar{\rho}_{B_{2g}}(|t'|=0.4)$ regardless of the sign of $t'$. Similarly due to the quadratic $t'$ dependence of ${\cal J}^2_{A_{1g}}$, $\bar{\rho}_{A_{1g}}(t'=0)>\bar{\rho}_{A_{1g}}(|t'|=0.2)>\bar{\rho}_{A_{1g}}(|t'|=0.4)$. 

Typically negative $t'$ leads to stronger correlation and suppresses the quasi-particle peak \cite{SP} and hence for a certain $|t'|$, $\bar{\rho}_{\alpha}(t'<0)>\bar{\rho}_{\alpha}(t'>0)$ is generally true except for the $A_{1g}$ case. In this exception, the negative linear $t'$ term in ${\cal J}^2_{A_{1g}}$ shifts the stationary point away from $t'=0$ and counters this effect from $\Upsilon(k,0)$ for small $|t'|$ leading to $\bar{\rho}_{A_{1g}}(t'=-0.2)<\bar{\rho}_{A_{1g}}(t'=0.2)$ and $\bar{\rho}_{A_{1g}}(t'=-0.4)\approx\bar{\rho}_{A_{1g}}(t'=0.4)$.

Besides, $\bar{\rho}_{A_{1g}}$ shows rather different $T$-dependent behaviors between electron-doped $t'\geq 0$ and hole-doped $t'<0$ cases. At negative $t'$, $\bar{\rho}_{A_{1g}}$ increases almost linearly with temperature. But at zero or positive $t'$, $\bar{\rho}_{A_{1g}}$ first increases sharply up to a certain temperature scale depending on $t'$ and then crosses over to a region where the growth rate becomes much smaller.

\section{ Finite $\omega$ results} \label{ACT}

Next we present the $\omega$-dependent optical and Raman conductivities defined in \disp{optical}. In  \figdisp{opsigmatpn2varyd} and \figdisp{opsigmatpp2varyd}, the set of four {$\omega$-dependent} conductivities are displayed for the hole-doped system at $t'=-0.2$ and the electron-doped system at $t'=0.2$ respectively for  a set of typical densities  at a low T. In the insets we display the corresponding imaginary part of  susceptibility, related through \disp{DC}. In most cases, the quasi-elastic peak gets suppressed and shifts to higher frequency when reducing the carrier concentration. The only exception is $\bar{\chi}''_{A_{1g}}$ at $t'=0.2$. Its quasi-elastic peaks are considerably smaller than other geometries due to the fluctuation in the specific vertex and they get higher and broader as doping increases. 

 In \figdisp{tpp2opsigmavaryT} we focus on the electron-doped case of varying $T$ at $t'=0.2$, $\delta=0.15$ where high quality experimental results are available for the $B_{2g}$ Raman channel in \refdisp{Koitzsch}, see particularly Fig.~(2). We evaluate the susceptibility at $T$ values  corresponding to those in this experiment. There is a fair  similarity between the theoretical curve (Panel (d)) and  the experiment. In particular, the theoretical curve reproduces  the quasi-elastic peak, and its T evolution. The other three panels in \figdisp{tpp2opsigmavaryT} are our theoretical predictions, and they are equally amenable to experimental verification. 
 
In the $xx$, $B_{1g}$, $B_{2g}$ geometries, the quasi-elastic peaks in susceptibility get slightly higher and quite broader upon warming. The $A_{1g}$ case is different. Its quasi-elastic peaks are much less obvious (too broad) except for the lowest temperature, and the peak magnitude is rather sensitive to temperature increase.

\begin{figure*}[!]
\subfigure[{Optical conductivity $\bar{\sigma}_{xx}$ and susceptibility $\bar{\chi}''_{xx}$}]{\includegraphics[width=.5\columnwidth]{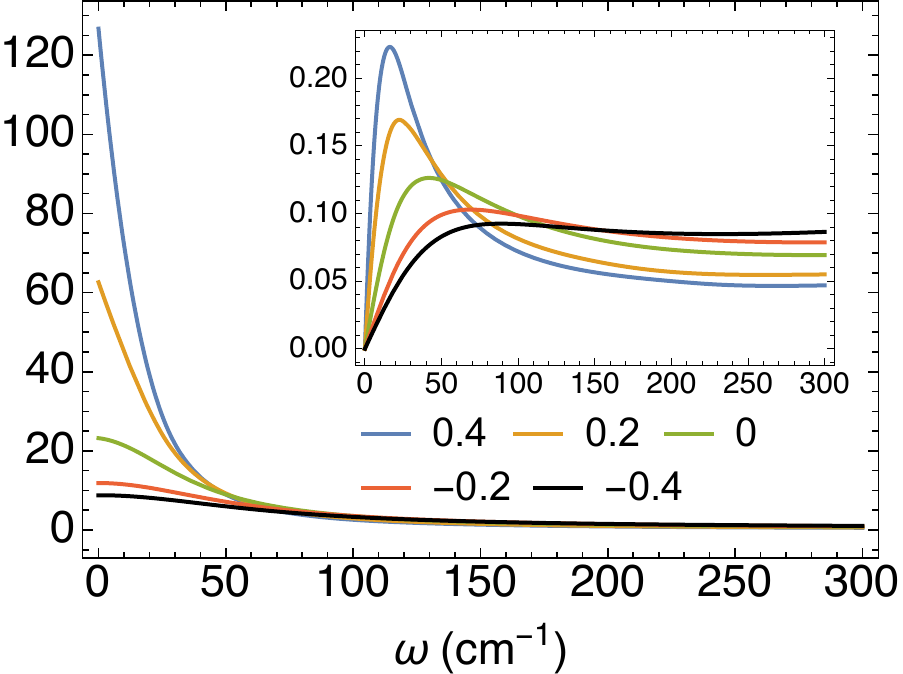}}
\subfigure[{$A_{1g}$ Raman: $\bar{\sigma}_{A_{1g}}$ and $\bar{\chi}''_{A_{1g}}$ (inset)}]{\includegraphics[width=.5\columnwidth]{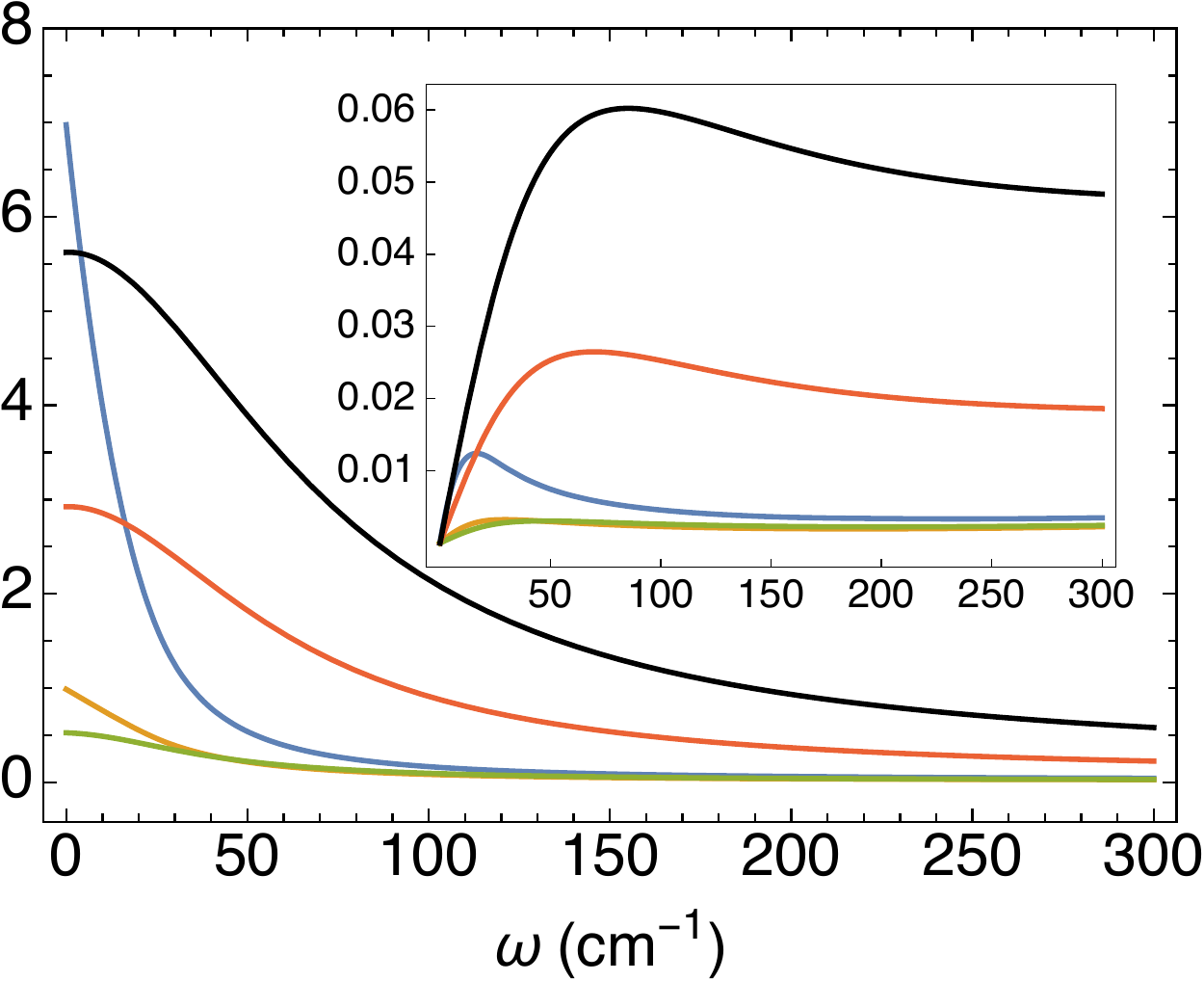}}
\subfigure[{$B_{1g}$ Raman: $\bar{\sigma}_{B_{1g}}$ and $\bar{\chi}''_{B_{1g}}$ (inset)}]{\includegraphics[width=.5\columnwidth]{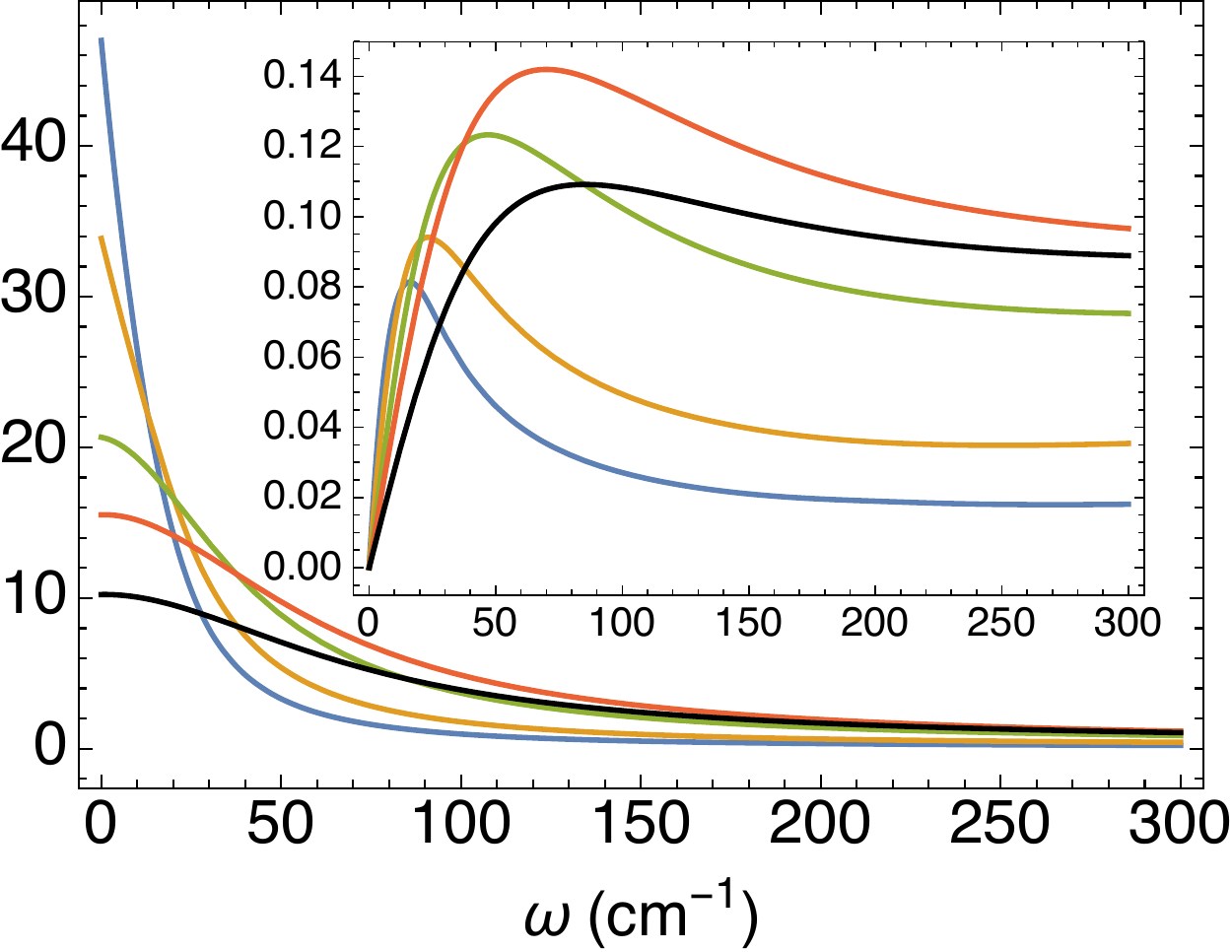}}
\subfigure[{$B_{2g}$ Raman: $\bar{\sigma}_{B_{2g}}$ and $\bar{\chi}''_{B_{2g}}$ (inset)}]{\includegraphics[width=.5\columnwidth]{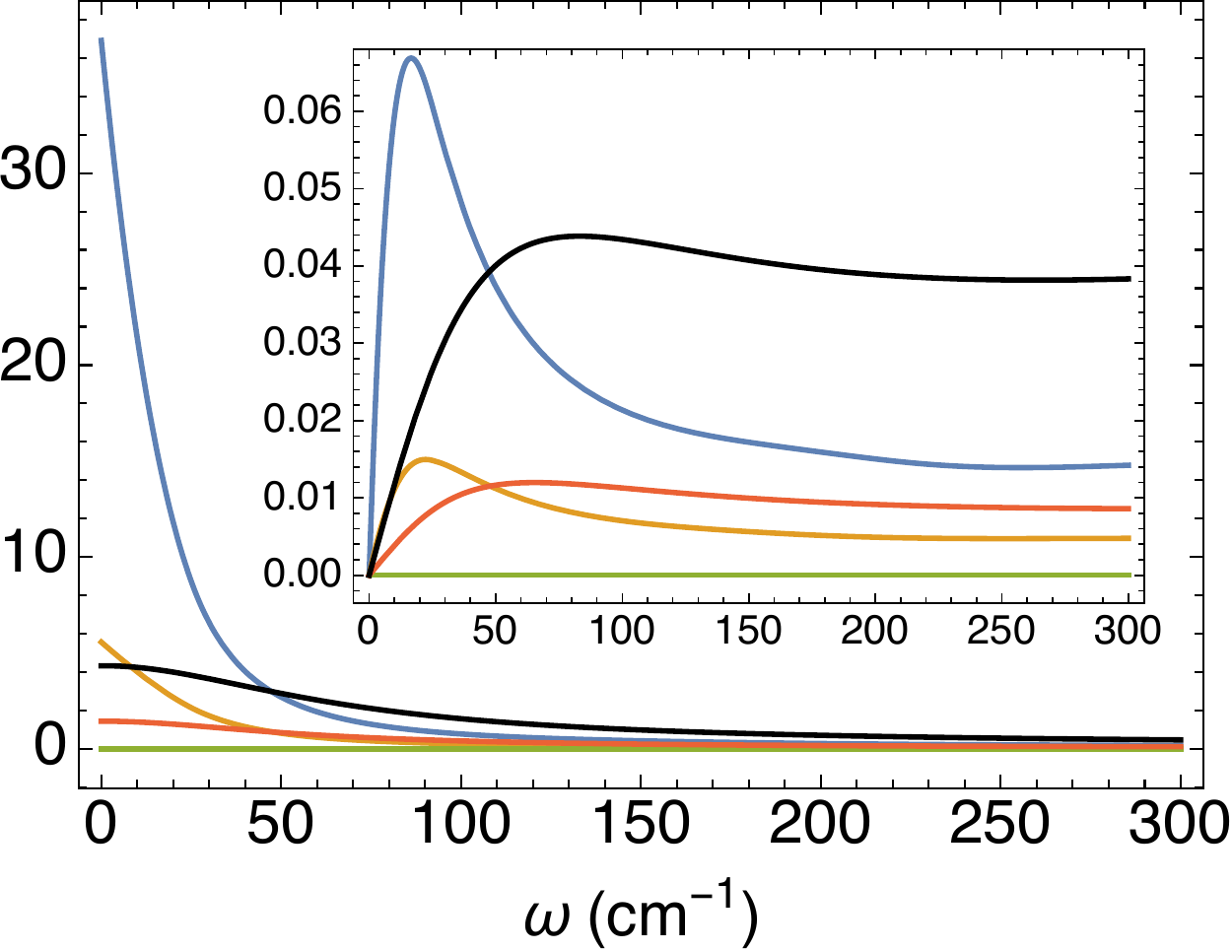}}
 \caption{ 
  \label{opsigmavarytp} Optical conductivity $\bar{\sigma}_{xx}(\omega)$ and the Raman conductivities $\bar{\sigma}_{A_{1g}}(\omega)$, $\bar{\sigma}_{B_{1g}}(\omega)$, $\bar{\sigma}_{B_{2g}}(\omega)$ at $\delta=0.15$, $T=63$K and varying $t'$, as marked (same legend for all subfigures). The corresponding dimensionless susceptibility is plotted in the inset with the same x-axis.  \refdisp{Sugai-2}, \refdisp{Sugai-3} and \refdisp{Koitzsch} show data that corresponds to these variables.}
 \end{figure*}

 \begin{figure*}[!]
\subfigure[{Optical Relaxation rate }]{\includegraphics[width=.49\columnwidth]{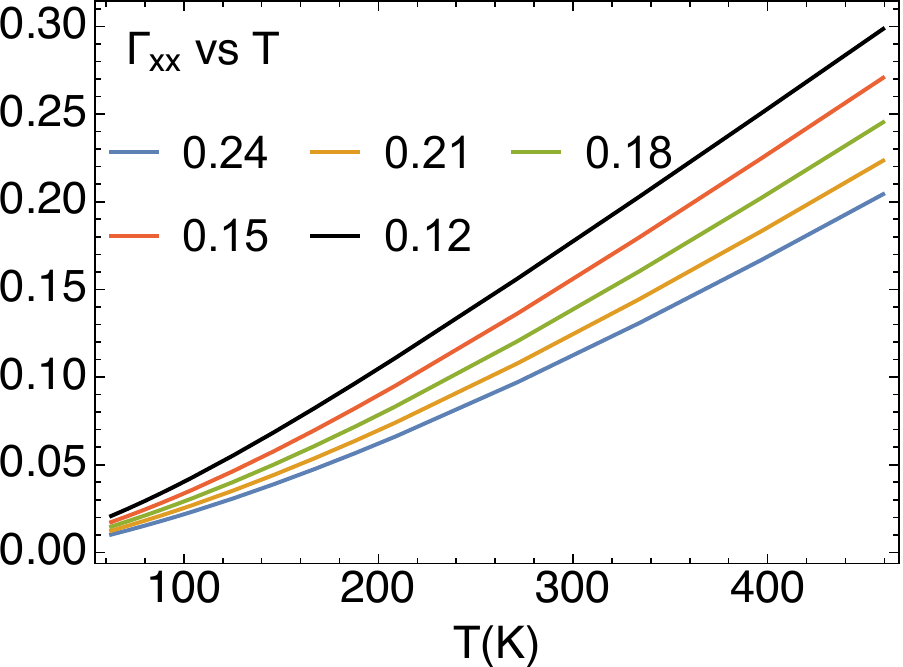}}
\subfigure[{$A_{1g}$ Raman: Relaxation rate  }]{\includegraphics[width=.49\columnwidth]{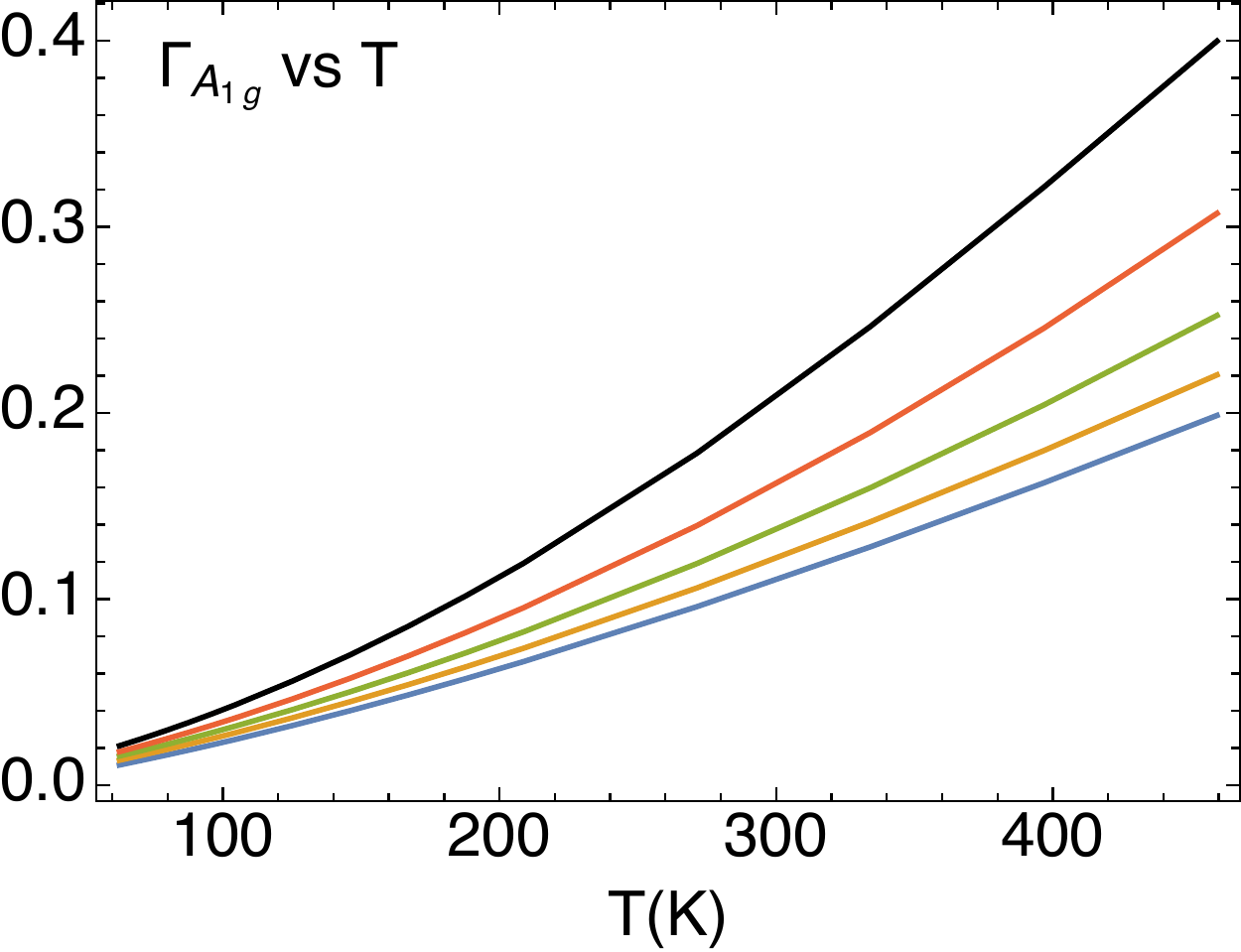}}
\subfigure[{$B_{1g}$ Raman: Relaxation rate  }]{\includegraphics[width=.49\columnwidth]{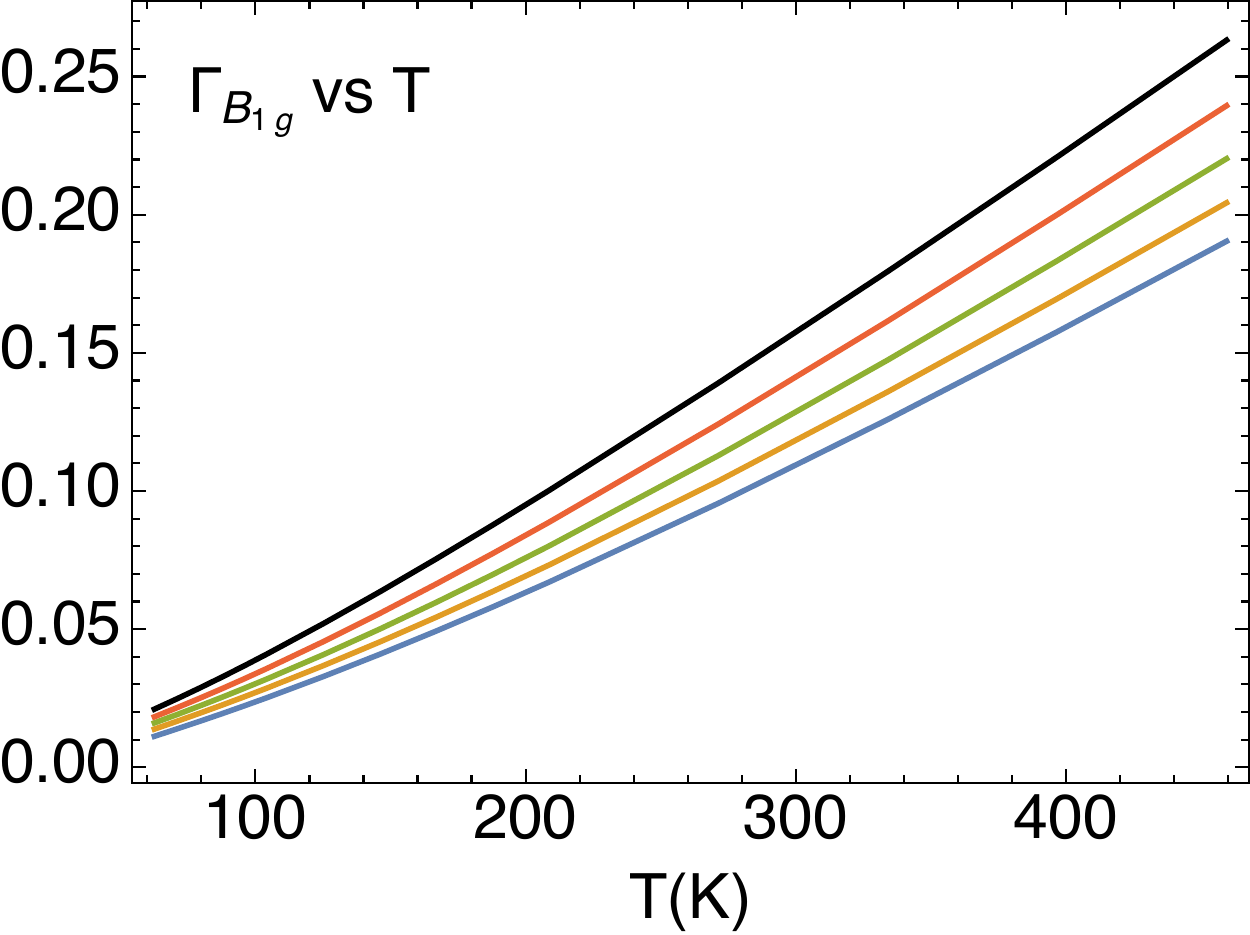}}
\subfigure[{$B_{2g}$ Raman: Relaxation rate  }]{\includegraphics[width=.49\columnwidth]{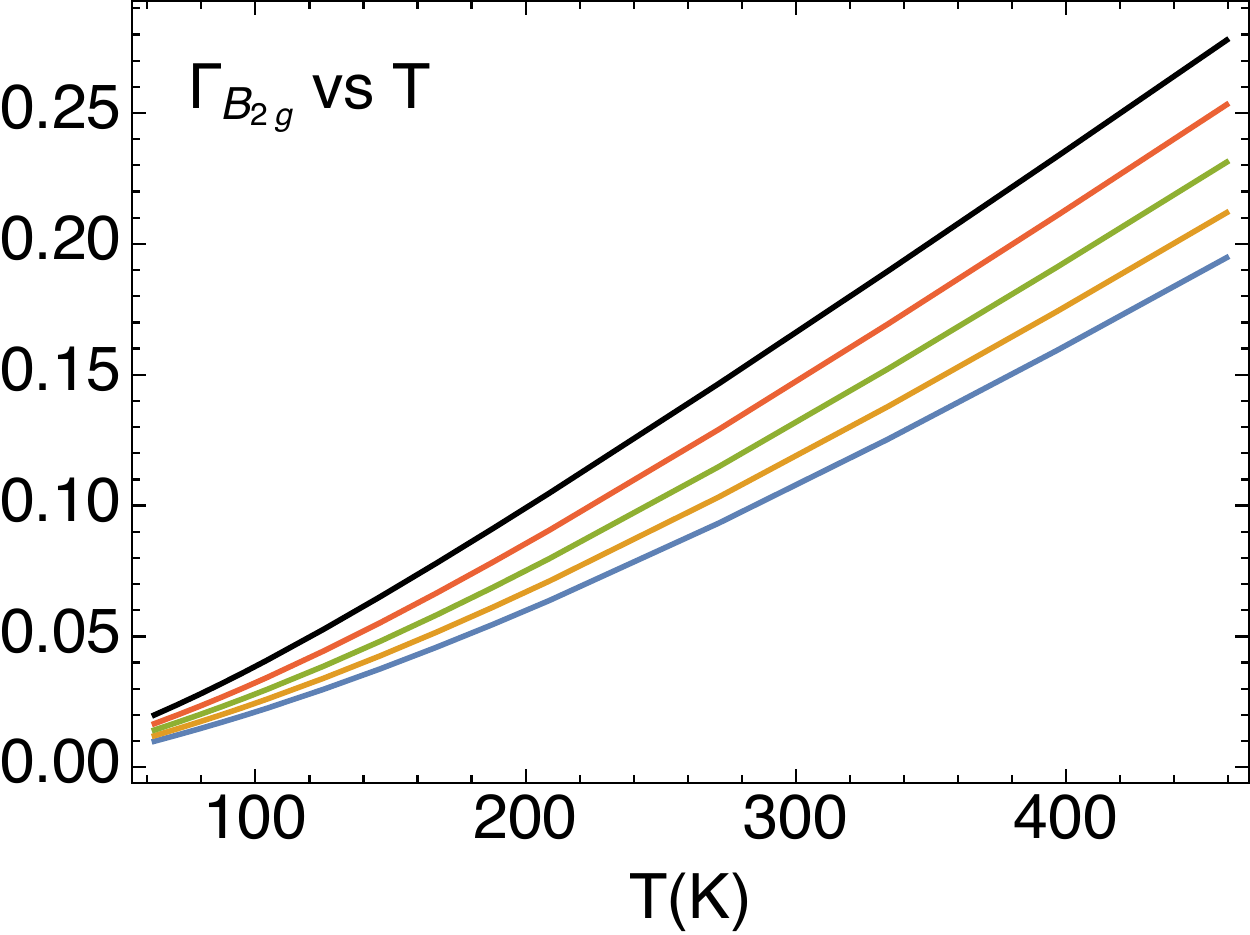}}
 \caption{ 
  \label{gammavarydelta} Relaxation rates (half widths at half maximum) of $\sigma_\alpha(\omega)$ in units of $t$, at $t'=-0.2$ at various marked $\delta$. The optical rate shows less convexity than the corresponding dc resistivity of \refdisp{SP}. The  rates in (a,b) and in (c,d) have similar  orders of magnitude, for reasons discussed in \figdisp{DCrhovaryd}. }
 \end{figure*}

\begin{figure*}[ht]
\subfigure[{Optical Relaxation rate }]{\includegraphics[width=.49\columnwidth]{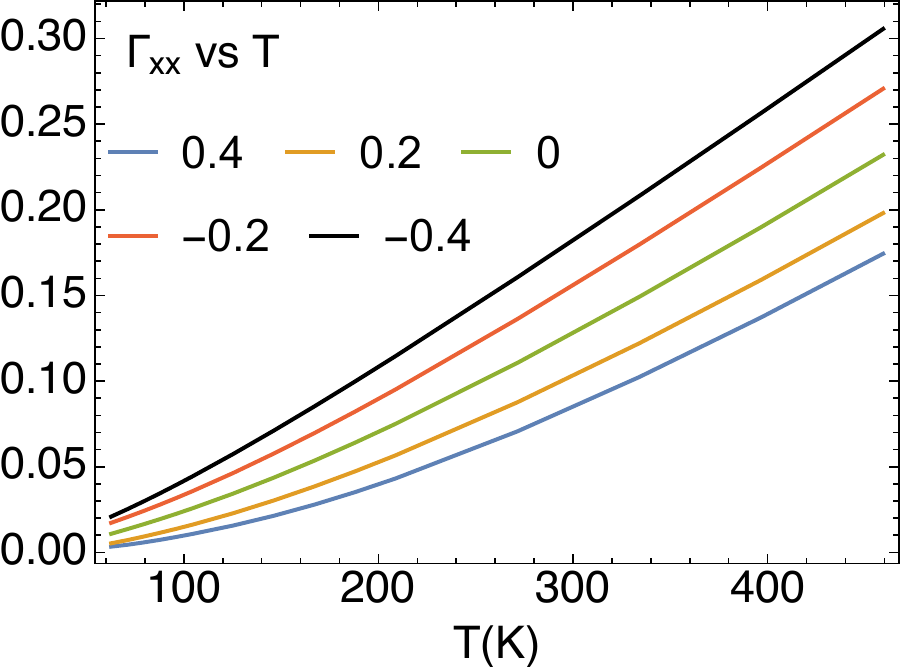}}
\subfigure[{$A_{1g}$ Raman: Relaxation rate}]{\includegraphics[width=.49\columnwidth]{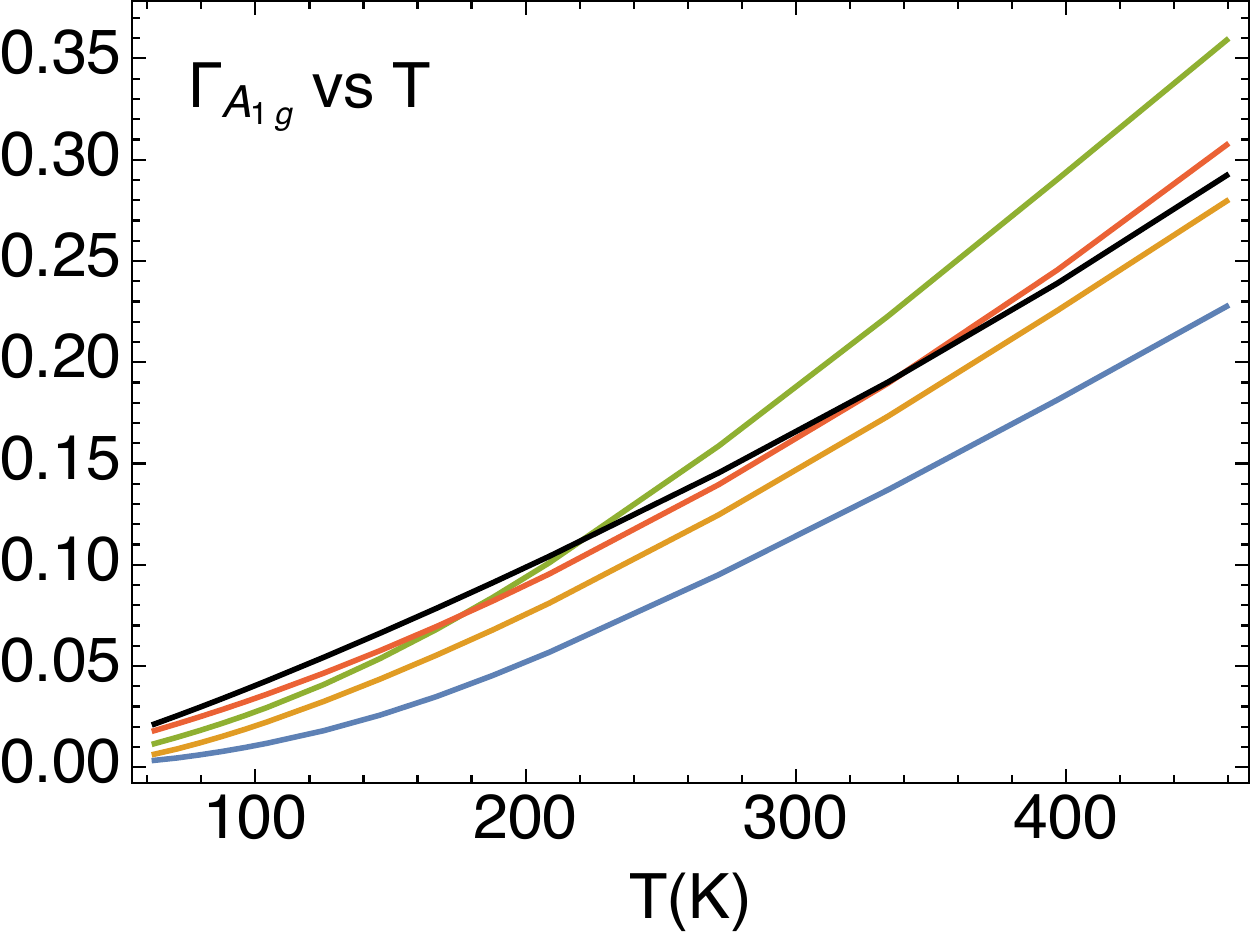}}
\subfigure[{$B_{1g}$ Raman: Relaxation rate}]{\includegraphics[width=.49\columnwidth]{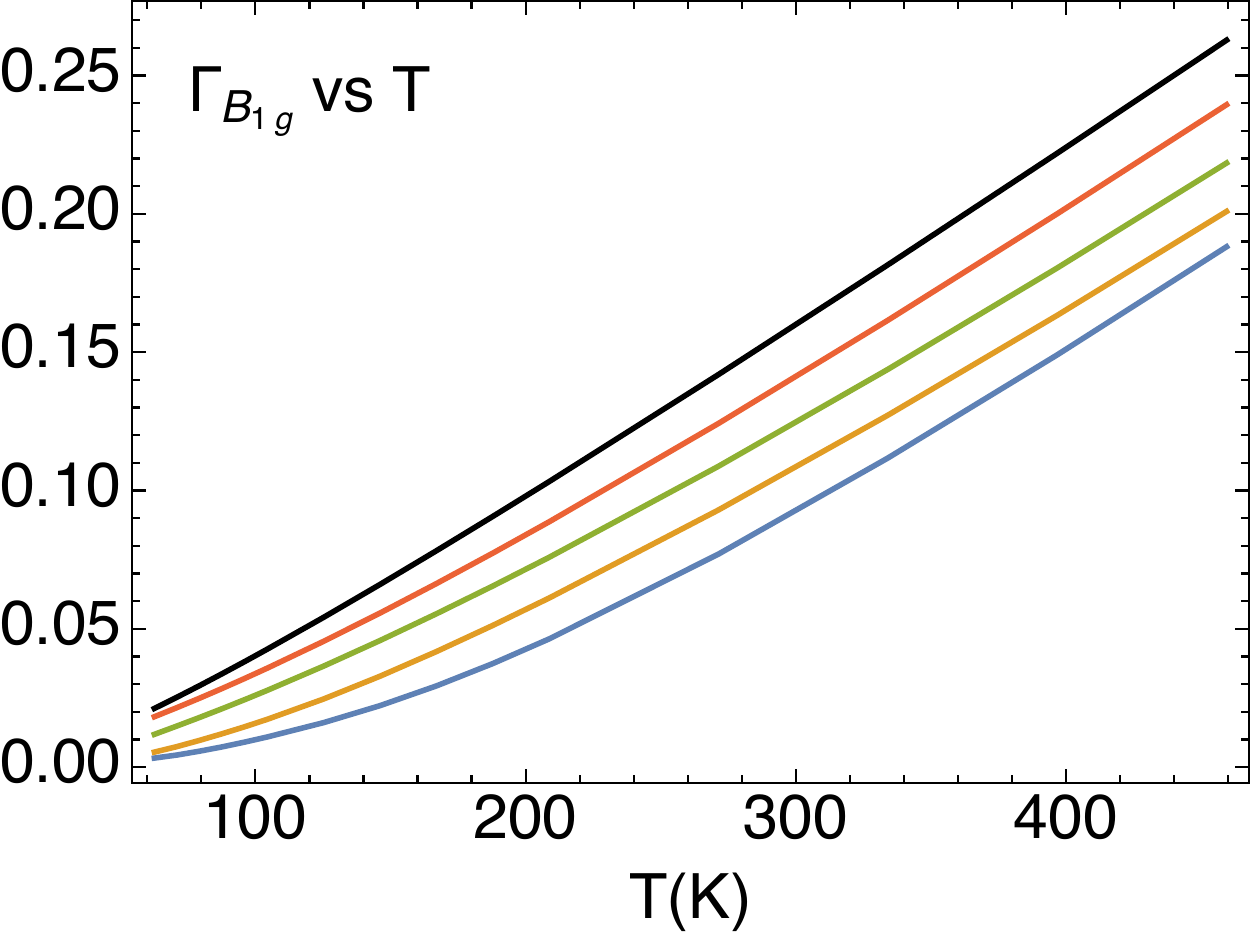}}
\subfigure[{$B_{2g}$ Raman: Relaxation rate}]{\includegraphics[width=.49\columnwidth]{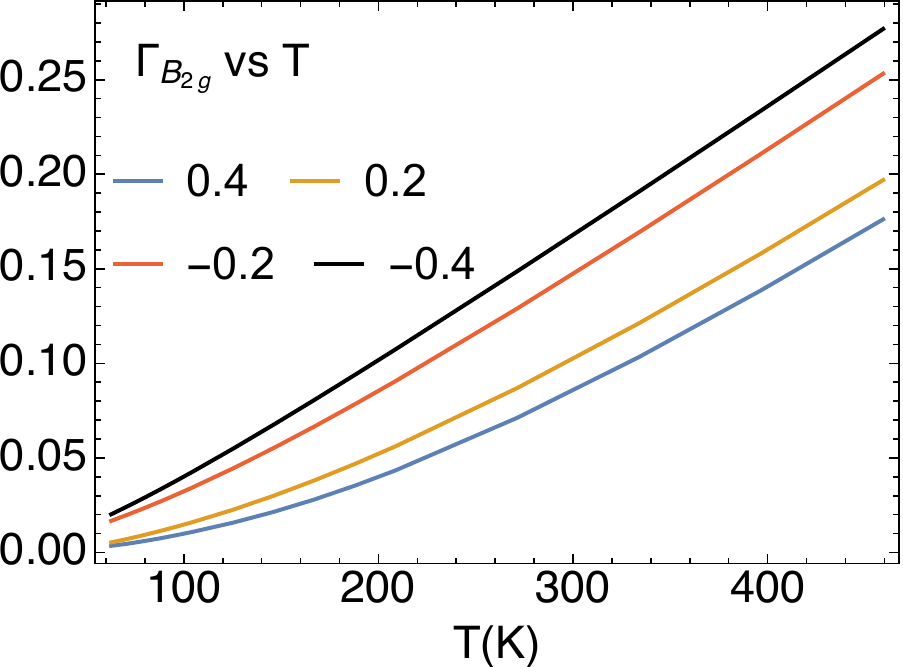}}
 \caption{ 
  \label{gammavarytp} The half-width at half-maximum for optical conductivity and Raman conductivities at $\delta=0.15$ and varying $t'$, as marked.}
 \end{figure*}

We also vary $T$ at hole doping $t'=-0.2$ in \figdisp{opsigmavaryT}. Comparing with the electron-doped case in \figdisp{tpp2opsigmavaryT}, we note that the hole-doped optical and Raman objects share a greater similarity in shape dependence on $T$, if we ignore the  scale difference. As T increases, the quasi-particle peaks get softened, and hence it generally suppresses the conductivities as well as the quasi-elastic peak in susceptibilities.  

For completeness, the $t'$ variation in  $\bar{\sigma}_{\alpha}(\omega)$ and $\bar{\chi}''_{\alpha}(\omega)$ is plotted in \figdisp{opsigmavarytp}, and it looks rather different among various geometries. This can be understood as arising from  the competition among various factors. We have a quadratic $t'$ dependence in the squared vertices, and a monotonic $t'$ dependence in the magnitude and geometry of $\Upsilon(k,\omega)$.  The $t'$ dependence of the shape of $\bar{\sigma}_\alpha$ has more commonality. Another interesting observation is that, unlike the DC case when $\bar{\sigma}_{xx}$ and $\bar{\sigma}_{B_{1g}}$ are similarly affected by $t'$, at  finite frequency, their behaviors depend on $t'$ rather differently. This difference is  more obviously observed in terms of $\bar{\chi}''$. 

From the optical and Raman conductivities $\bar{\sigma}_{\alpha}$  we can extract a frequency scale $\Gamma_{\alpha}$, as the  half-width at half-maximum, in the unit of $t$. These are plotted against T in Fig.~(\ref{gammavarydelta}) for varying $\delta$ and \figdisp{gammavarytp} for varying $t'$. It is remarkable that despite a bare band width of $\sim 3.6$ eV, these frequency scales appear close to linear in T down to very low T. This is closely related to the observation in \refdisp{SP} that the resistivity departs from a $T^2$ behavior at extraordinarily low $T$'s, i.e. the effective Fermi temperatures are suppressed from the bare values by two or more orders of magnitude. Although the magnitude of the optical and Raman conductivities differs a lot, their relaxation rates describing the shape turn out to be much closer, as a result of a similar $T$-dependent line shape of the spectral function\cite{SP} in the normal state.

\section{ Conclusion and Discussion}

We have presented calculations of  the electrical and Raman resistivities in the dc limit, the optical conductivity, the Raman susceptibilities and related objects based on the second order ECFL theory in \refdisp{SP}.  We computed the susceptibilities (using the leading order approximation) with the shown results. Experiments on different geometries can test and put some bound on  this hypothesis of weak vertex corrections for the Raman operators. This is clearly of theoretical  importance, since going beyond the bubble graphs brings in a formidable level of complexity.

The ECFL theory leads to a very small quasi-particle weight $Z$ and a large background extending over the bandwidth, and it has a very small effective Fermi temperature leading to an interesting $T$ dependence of the resistivity as discussed in \cite{SP}.  The line shape of the calculated Raman susceptibility is  close to that  for the case of electron doped NCCO \refdisp{Koitzsch} in terms of $T$ and $\omega$ dependence, and therefore is  promising. Our calculation also gives the Raman susceptibility in two other geometries, and this prediction can be checked against future experiments that are quite feasible. We note that  the data \refdisp{Sugai-2} from Sugai {\em et. al.}  for this quartet of variables in the case of LSCO seems roughly consistent with our  results, and a more detailed comparison is planned.

The focus on the $T$ dependence in the $\omega\to 0$ limit, i.e. on  resistivities can be quite a fruitful goal for  future experiments, since this  limit gets rid of all excitations and measures  the ``pure-background''. It is an important exercise since the different geometries probe different combinations of $t,t'$ as they occur in the bare vertices \disp{vertices}, as stressed above. We are predicting that the Raman resistivity in each channel can be found from the intensity at low $T$, and broadly speaking similar to resistivity. In further detail, it is predicted to be (a) channel specific and (b) $t'/t$ dependent. These  clearcut predictions can be tested in future experiments.

Finally although such a measurement  is not  commonly done, a systematic  measurement of   the ratios of the scattering cross sections in different geometries  should be feasible. This measurement, and a comparison between the quartet of susceptibilities presented here,  can be profitably compared  with recent  theories of strongly correlated systems to yield material parameters. Most importantly it can yield  physical insights into the mechanism underlying the  broad  nonresonant Raman signals that have remained  quite mysterious  so far.

\section{ Acknowledgement:} 
We thank Tom Devereaux,  Lance Cooper and Girsh Blumberg for helpful discussions.
The computation was done on the comet in XSEDE \cite{xsede} (TG-DMR170044) supported by National Science Foundation grant number ACI-1053575. The work at UCSC was supported by the US Department of Energy (DOE), Office of Science, Basic Energy Sciences, under Award No. DE-FG02-06ER46319.


\begin{thebibliography}{99}
\bibitem{Comment-A1g} Parabolic bands leads to an exact cancellation, whereby the  $A_{1g}$ scattering  is  unobservable. References. ~ \cite{Wolff-1,Abrikosov-Genkin,raman1,raman2}  show that the cancellation is inoperative with non-parabolic bands, or with disorder. Experimentally, Fig.~(2) of \refdisp{Sugai-2} shows an  $A_{1g}$ contribution   similar in scale  to the other two geometries.
\end{thebibliography}
\end{document}